\documentclass[aps, twocolumn, superscriptaddress]{revtex4-1} 
\usepackage{amssymb, amsmath, amsthm}
\usepackage{xcolor}
\usepackage{graphicx}
\usepackage[protrusion=true,expansion=true]{microtype}
\usepackage{times}
\usepackage{hyperref}
\hypersetup{
    pdftitle={main},    
    pdfauthor={},
    colorlinks=true,
    citecolor=blue,
    linkcolor=blue,
    urlcolor=blue
  }
\usepackage{comment}
\usepackage{blkarray}
\usepackage{mathtools}
\usepackage{esint}
\usepackage{tikz}
\usetikzlibrary{decorations.pathreplacing}
\usepackage{dsfont}
\usepackage{braket}

\usepackage[normalem]{ulem}
\usepackage{graphicx}
\usepackage{dcolumn}
\usepackage{bm}

\newcommand{\prlsection}[1]{{\em {#1}.---~}}
\newcommand{\Tr}[1]{\mathrm{Tr}}

\newcommand{\ii}{{\rm i}}

\renewcommand{\vec}[1]{\boldsymbol #1}

\def\nn{\nonumber}
\newcommand{\dr}[1]{{#1}^{\rm dr}}

\usepackage{bm}
\newcommand{\be}{\begin{equation}}
\newcommand{\ee}{\end{equation}}
\newcommand{\de}{\partial}

\newcommand{\Ai}{\text{Ai}}

\def\Tr{\operatorname{Tr}}
\renewcommand{\vec}{\mathbf}

\begin{document}

\title{Quantum fluctuating theory for one-dimensional shock waves}

\author{Andrew Urilyon}
\affiliation{Laboratoire de Physique Th\'eorique et Mod\'elisation, CNRS UMR 8089,
CY Cergy Paris Universit\'e, 95302 Cergy-Pontoise Cedex, France}

\author{Stefano Scopa}
\affiliation{Laboratoire de Physique Th\'eorique et Mod\'elisation, CNRS UMR 8089,
CY Cergy Paris Universit\'e, 95302 Cergy-Pontoise Cedex, France}

\author{Giuseppe Del Vecchio Del Vecchio}
\affiliation{Universit\'e Paris-Saclay, CNRS, LPTMS, 91405, Orsay, France}

\author{Jacopo De Nardis}
\affiliation{Laboratoire de Physique Th\'eorique et Mod\'elisation, CNRS UMR 8089,
CY Cergy Paris Universit\'e, 95302 Cergy-Pontoise Cedex, France}

\begin{abstract}
We study the formation and the subsequent dynamics of shock waves in repulsive one-dimensional Bose gases during the free expansion of a density hump. By building coherent Fermi states for interacting Bethe fermions, we define a quantum fluctuating initial state expressed in terms of universal quantities, namely the density and the Luttinger parameter. In the integrable case, this fluctuating state is then evolved by generalized hydrodynamics (GHD) and, differently from non-fluctuating initial states, it develops density ripples on top of the hydrodynamic mean value. Our analysis gives a general theory of quantum ripples and wave breaking in integrable and quasi-integrable one-dimensional liquids and clarifies the role of the interaction strength. In particular, for strongly/intermediately interacting bosons, we find quantum ripples originating from low-energy modes at the Fermi surface interfering when transported by GHD. In the low coupling limit, near the quasicondensate regime, we find instead that density ripples have a semi-classical nature, and their description  requires information on the curvature of the Fermi surface.
\end{abstract}

\maketitle

\section{Introduction}\label{sec:intro}
\begin{figure}[t]
\centering
  \includegraphics[width=0.4\textwidth]{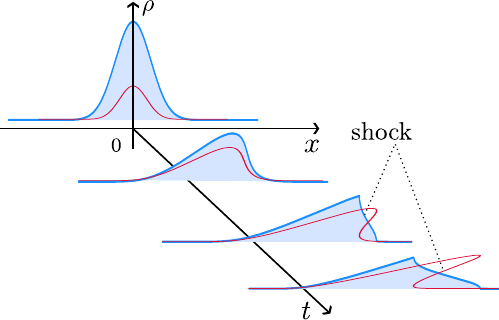}
  \caption{Illustration of a shock wave forming from a density perturbation in one-dimensional quantum gases. Red curves depict the corresponding evolution of velocities (or rapidity for interacting integrable systems): when fast modes overturn the background a shock wave is formed.
  }\label{fig:shock}
\end{figure}

Quantum gases in one dimension often respond to external perturbations through collective excitations rather than single-particle effects. An illustrative perturbation consists of a spatial modulation of the fluid density, induced e.g. by an external potential (Figure~\ref{fig:shock}), with Tomonaga-Luttinger liquid theory \cite{Giamarchi2003,Haldane1981,Haldane1981a,Imambekov2012} providing a universal method for small amplitude perturbations in linear response. Beyond linear response, a fascinating class of phenomena emerges. A celebrated example is the formation of shock waves in a fluid of impenetrable bosons (or noninteracting fermions). Here, denser regions have higher particle velocities, causing an overturn of fast modes over the background, and initiating nonlinear physics after the shock time $t_s\sim \Delta x/\Delta\rho$, see Fig.~\ref{fig:shock}.
 Shock waves in free Fermi gases have been have been discussed in many works,  e.g.~\cite{Balazs1973,damski2004,Bettelheim2006,Bettelheim2007,Bettelheim2008,kulkarni2012,Bettelheim2012,PhysRevA.80.043606,Protopopov2013,PhysRevResearch.3.013098,PhysRevA.104.023316}, and more recently studied in interacting integrable models through the use of Generalized Hydrodynamics (GHD) \cite{Scopa2022b,Scopa2021real,urichuk2023navierstokes,Doyon2017,Dubessy2020,simmons2023,Moller2024}, often together with Gross-Pitaevskii equation, truncated Wigner approach and different numerical methods \cite{Doyon2017,Simmons2020,simmons2022phase,simmons2023,Bettelheim2019,Dubessy2020,Watson2024, Moller2024}. However a full characterization, especially in presence of non-trivial interactions, is far from being complete.  For integrable, or quasi-integrable, gases, GHD, and in particular its zero-temperature limit (zero-entropy GHD), gives an accurate mean description of the pre and post-shock dynamics, but it fails to capture quantum fluctuations, particularly in scenarios like the one in Figure~\ref{fig:shock}. Indeed, at low temperatures, quantum fluctuations play an important role in the dynamics, as with dynamics the so-called quantum ripples are transported on large scales, giving this way macroscopic wave effects.  Therefore they need to be included for a complete hydrodynamic description. One way to do it is by the so-called quantum GHD e.g.~\cite{Ruggiero2020,Scopa2021,Scopa2022,Scopa2023,Collura2020,Ruggiero2021,takacs2024quasicondensation}, where linearized modes around the zero-entropy GHD background are quantized in terms of Tomonaga-Luttinger chiral bosons, and evolved according to an effective quadratic Hamiltonian. Although intriguing, this method encounters some technical challenges, primarily in handling the evolution of bosonic correlations within the GHD framework.\\
In this work, we take a different path to merge quantum fluctuations with hydrodynamic descriptions, in particular focusing on the case of shock wave dynamics. Analogously to what is done in classical \textit{fluctuating hydrodynamics}, we construct an initial state containing quantum fluctuations, and we use it to build a fluctuating phase-space occupation function $n_\text{Fluct}(x,k)$ that can be evolved with the standard GHD equations of Refs.~\cite{Castro-Alvaredo2016,Bertini2016}, or once provided a Boltzmann-like equation for the non-integrable case \cite{panfil2023thermalization,bertini2015pretherm,durnin2021nonequilibrium,piqueres2023integrability,essler2014quench,delvecchiodv2022,bastianello2021hydrodynamics}. Expectation values of local densities are then obtained as weighted integrals in quasimomentum space, displaying fluctuations around their mean value. This approach allows us to give a complete characterization of quantum and semi-classical density ripples during the formation and propagation of the shock, thereby establishing a direct relation between the types of ripples and the underlying microscopic interaction in the gas. \\

In general, characterizing such fluctuating initial states for an interacting quantum gas is a hard task. In the case of Fig.~\ref{fig:shock} and restricting to noninteracting fermions, the system is found in a coherent Fermi state \cite{Bettelheim2011,Keeling2006,Keeling2008,Ivanov1997,Perelomov1986}, a superposition of particle-hole excitations generated by the external potential on top of a Fermi sea. Here we extend this description of coherent states of fermions to interacting quantum gases, such as the interacting Lieb-Liniger model. Namely, we construct fluctuating initial states obtained as coherent states of the refermionized degrees of freedom describing quantum fluctuations at low energies. In particular, for a given initial macroscopic density pertubation $\rho(x)$, we find a density of Bethe fermions $\rho_F(x)=\rho(x)/\sqrt{K(x)}$, with $K(x)$ the local Luttinger parameter describing the emergent one-dimensional quantum fluid \cite{Giamarchi2003,Dubail2017}. The resulting \emph{fermionic occupation function}, $n_\text{Fluct}(x,k)$ is universal, being  dependent only on the density $\rho(x)$ and on the Luttinger parameter $K(x)$. In the integrable case, its time evolution is carried out with GHD, which proved its efficacy in numerous studies e.g.~\cite{bulchandani2017solvable,Bulchandani2018,Gopalakrishnan2018,piroli2017transport,caux2019hydrodynamics,Bastianello2019,DeLuca2017,Collura2018,Koch2021adiabatic,bastianello2020generalized}, 
including different experimental tests \cite{tang2018,Schemmer2019,Malvania_2021,moller2021extension,scheie_detection_2021,jepsen_spin_2020,Le2023,yang2023phantom,cataldini2022emergent,2312.15344}. In the non-integrable case, one can consider the kinetic picture for the refermionized quasiparticles proposed in Ref.~\cite{protopopov2014}.

\begin{figure}[t]
\centering
  \includegraphics[width=0.4\textwidth]{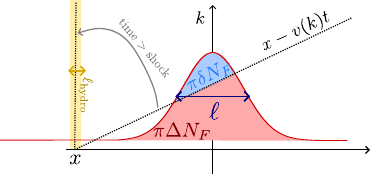}
  \caption{Illustration of the quantum fluctuations generating around position $x$ at time $t>t_s$ after the shock formation. The number of Bethe fermions $\delta N_F(x)$ undergoing quantum interference is obtained as the area on the initial density hump between those points that are transported by Euler hydrodyamics to position $x$, and it must be compared to the total number of Bethe fermions $\Delta N_F$ in the hump.
  }\label{fig:fig2}
\end{figure}
Our analysis finally clarifies the different nature of density ripples in one-dimensional shock waves depending on the interaction coupling of the gas. 
To clarify this, we focus on the post-shock dynamics $t>t_s$ and we consider a number of Bethe fermions $\delta N_F(x)$ contained within a small subregion of size $\delta x\sim {\cal O}(\ell_\text{hydro})$ around position $x$, see Fig.~\ref{fig:fig2} for an illustration. Here, we introduced the hydrodynamic scale $\ell_\text{hydro}$ satisfying $\ell_\text{hydro} \ll \ell$, with $\ell\sim \rho/|\de_x\rho|$ the scale of the external perturbation. A quantization condition in this subregion requires that momenta satisfy $(\delta k \cdot \delta x)/\Delta N_F= \delta N_F/\Delta N_F$, with $\Delta N_F$ the total number of particles in the hump. We then distinguish the following cases: \\
\indent-- \emph{Quantum regime}, if $\delta N_F/\Delta N_F\sim{\cal O}(1)$. Quantum ripples are associated to the notion of a quantized phase space, and are generated from the interference of low-energy modes at the Fermi surface. This class includes those effects that are captured by field-theory approaches, such as quantum GHD, e.g.~\cite{Ruggiero2020,Ruggiero2021,Scopa2023}.\\
\indent-- \emph{Semi-classical regime}, if $\delta N_F/\Delta N_F\ll 1$. In this case, to properly characterize the density ripples around position $x$, further information on the Fermi surface in a surrounding region scaling as $\sim {\cal O}(\ell^{2/3})$ becomes necessary. Such semi-classical ripples might be viewed as a background effect originating from higher-derivative corrections \cite{Fagotti2017,Fagotti2020,denardis_doyon_dispersive,Bocini2023}, and/or by universal corrections in the form of Airy kernels at the propagating front, e.g.~\cite{Balazs1973,Bettelheim2011,Bettelheim2012,Eisler2013Full,Dean2018Wigner,Dean2019,DeBruyne2021wigner,gouraud2024quantum}. \\
Both quantum and semi-classical ripples are captured by our fluctuating state, with the dominant contribution fixed by the choice of the initial density hump $\Delta\rho$ and the value of the coupling. At strong repulsion $c\to\infty$, Bethe fermions coincide with physical degrees of freedom ($K= 1$), thus one can choose a density perturbation such that $\delta N(x)\sim{\cal O}(\Delta N)$, and quantum ripples give the leading contribution inside the shock region. Moreover, since the gas is non homogeneous, semi-classical ripples are also present, particularly in regions $\sim{\cal O}(\ell^{2/3})$ around the shock fronts \cite{Bettelheim2011,Bettelheim2012}. In the weak coupling regime $c\to 0^+$ and $\rho\sim 1/c\sim{\cal O}(\ell)$ to the quasicondensate, the density ripples only exhibit a semi-classical nature. In this limit, the Luttinger parameter diverges as $K\sim 1/\sqrt{c}$ \cite{popov_theory_1977}, leading to a diverging Bethe fermion density $\rho_F\sim 1/\sqrt{c}$, with semi-classical ripples spreading over a size ${\cal O}(c^{-2/3})$.  Finally, for intermediate couplings both quantum and semi-classical ripples are present with the quantum ones rapidly vanishing as interactions are decreased.\\

\prlsection{Outline}
The paper is organized as follows. In Sec.~\ref{sec:model}, we introduce the Lieb-Liniger model, its Bethe ansatz solution, and the description of its low-energy universal excitations by refermionization methods. In Sec.~\ref{sec:tonks}, we revisit the case at strong repulsion (after Ref.~\cite{Bettelheim2011}): we build the fluctuating initial state, and we evolve it with Euler hydrodynamics. We then provide analytical expressions for the density ripples (both quantum and semi-classical). Sec.~\ref{sec:interact-method} contains our argument to construct fluctuating initial states at finite interactions. The semi-classical limit of Lieb-Liniger to the quasicondensate regime is considered in Sec.~\ref{section:semi-classical_limit}, unveiling the importance of higher-order corrections to the saddle point contribution at small coupling. Sec.~\ref{sec:discuss} discusses our results in comparison with numerical simulations. Finally, Sec.\ref{sec:conclusion} contains our conclusions and some perspectives. Some technicalities and some further discussions are deferred to the Appendixes \ref{app:bosonization}-\ref{app:qghd-nls}.

\section{Model and refermionization of low-energy excitations}\label{sec:model}

Our analysis takes the Lieb-Liniger model \cite{Lieb1963,Lieb1963a} as a foundational reference, although our derivation maintains full generality. Describing $N$ contact-interacting bosonic particles subject to an external potential $V(x)$, the Lieb-Liniger model's Hamiltonian is
\begin{equation}\label{eq:lieb-liniger-H}
\hat{H} = -\frac{1}{2} \sum_{i=1}^N \frac{\de^2}{\partial {x_i}^2} + c \sum_{i<j=1}^N \delta(x_i-x_j) + \sum_{i=1}^N V(x_i),
\end{equation}
with the local density operator $\hat{\rho}(x) = \sum_j \delta(x-x_j)/L$ and with $L$ the system size. Below we specify to the repulsive regime $c>0$ and we shall use $\gamma=c/\rho$ as the rescaled coupling strength. This model serves as a paradigmatic representation for one-dimensional interacting systems and cold atomic gases, as extensively discussed in Refs.~\cite{Cazalilla2004,Cazalilla2011,Bouchoule2022}. \\
In the absence of external potentials (or equivalently when $V(x)=$ const), the Hamiltonian \eqref{eq:lieb-liniger-H} is integrable. Namely, for even (resp.~odd) $N$, an eigenstate of $\hat{H}$ is associated to a sequence of half-integers (resp. integers)  $\{I_i\}_{i=1}^N$ that specifies the set quasi-momenta $\theta_i \in [-\infty,\infty]$ (or rapidities) through the solution of the Bethe equations
\begin{equation}\label{eq:BetheEq}
    \theta_j = \frac{2 \pi I_j}{L} +  \sum_{k=1}^N \phi(\theta_j - \theta_k)
\end{equation}
with scattering phase $\phi(\theta) =  2\arctan(\theta/c)$ and the associated scattering shift $T = \partial_\theta \phi$. Physically, one can interpret $2\pi I_j/L$ as the momenta of $N$ non-interacting fermions, and the Bethe equations as a non-trivial quantization condition accounting for the interactions. {The system's eigenstates can be parameterized by either the integers or the rapidities, by application of the mapping} from $\ket{\{I_i\}_{i=1}^N}$ to $\ket{\{\theta_i\}_{i=1}^N}$ given in eq.~\eqref{eq:BetheEq}. For instance, the ground state set of rapidities $\{\theta_i\}_{i=1}^N$ is obtained by solving \eqref{eq:BetheEq} from an equally spaced sequence of Bethe integers $\{I_i\equiv i-\frac{N-1}{2}\}_{i=1}^N$.\\

In the thermodynamic limit $N,L \to \infty$, the set of rapidities densely populate the real line and it is conveniently described in terms of a density distribution. Precisely, macrostates of the Lieb-Liniger gas are described in terms of a Fermi-Dirac occupation function $n(\theta)$ (usually called filling function) for fermionic quasiparticles featuring a bare energy $\varepsilon(\theta) = \theta^2/2$ and bare momentum $p(\theta)= \theta$. Its precise expression is dictated by the value of temperature and chemical potential, see e.g. Ref.~\cite{takahashi1999} for a detailed discussion. At zero temperature, the filling function becomes a perfect Fermi sea between the two Fermi edges $\pm \theta_{F}$, namely $n(\theta)=\Theta(\theta^2-\theta_F^2)$ where $\Theta(\cdot)$ is the Heaviside step function. Expectation values of local observables are then obtained as integrals in rapidity space weighted with $n(\theta)$. The density is $\rho_0=\int \frac{d\theta}{2\pi} 1_{[n]}^{\rm dr}(\theta) n(\theta)$,  with  dressing operation respect to filling $n$,  defined as  
\be\label{eq:dr-operation}
f_{[n]}^{\rm dr}(\theta)=f(\theta)+\int \frac{d\alpha}{2\pi} T(\theta-\alpha) n(\alpha) f_{[n]}^{\rm dr}(\alpha),
\ee
and similarly for other quantities.
\\

As $n(\theta)$ describes a perfect Fermi sea, any information on ground state fluctuations is lost. A standard way to reintroduce these quantum fluctuations is offered by refermionization methods. For instance, quantum fluctuations of the density operator $\delta\hat\rho(x)=\hat\rho(x)-\rho_0$, can be written as
\be\label{eq:density-exp}
\delta\hat\rho(x)=\sqrt{K}(\hat\psi^\dagger_+ (x)\hat\psi_+(x)+\hat\psi_-^\dagger(x)\hat\psi_- (x)) +\text{h.o.c.}
\ee
where $\hat\psi_{\pm}$ are chiral fermions whose properties are established by a free massless Dirac field theory, see e.g. Refs.~\cite{Senechal1999,Gogolin1998} for further discussion. The two-point function is
\be
\langle \hat\psi^\dagger_\sigma(x)\hat\psi_{\sigma'}(y)\rangle=\frac{\delta_{\sigma,\sigma'}}{\ii\sigma(x-y)}, \quad \sigma,\sigma'=\pm
\ee
and higher correlations can be obtained using Wick's theorem. Alternatively, one can encode low-energy fluctuations in terms of Tomonaga-Luttinger chiral bosons $\hat\varphi_\pm$ satisfying $\langle\hat\varphi_\sigma(x)\hat\varphi_{\sigma'}(y)\rangle= \delta_{\sigma,\sigma'} \log(\ii\sigma(x-y))$. The two descriptions are connected via bosonization, 
$\hat\psi_\sigma(x) \propto :e^{-\ii\sigma\hat\varphi_\sigma(x)} :$
(where $:\cdot :$ is the normal ordering of fields), see Appendix~\ref{app:bosonization} for more details.
\\

Eq.~\eqref{eq:density-exp} can be also understood on physical grounds. Low-energy states of the Lieb-Liniger model can be obtained by creating particle-hole pairs around the left/right Fermi edges, and can be realized by left/right fermionic operators $\hat{c}^\dagger_{p\lessgtr 0}=\int dx \ e^{-\frac{\ii 2\pi px}{L}} \hat\psi^\dagger_\mp(x)$ acting on the sequence of ground state Bethe integers. For instance,
\be
\hat{c}^\dagger_{\frac{N}{2}}\hat{c}_{\frac{N+1}{2}}   \ket{\{\dots\underset{\frac{N-1}{2}}{\bullet}\underset{\frac{N+1}{2}}{\bullet}\underset{\frac{N}{2}}{\circ}\} } = \ket{\{\dots\underset{\frac{N-1}{2}}{\bullet}\underset{\frac{N+1}{2}}{\circ}\underset{\frac{N}{2}}{\bullet}\} } .
\ee

Hence, Eq.~\eqref{eq:density-exp} encodes the quantum fluctuations of density as coming from particle-hole pairs generated at the Fermi edges. The amplitude in eq.~\eqref{eq:density-exp} can be in fact interpreted as the thermodynamic form factor $\langle\text{part.-hole}|\delta\hat\rho(0)|\text{g.s.}\rangle/L=\sqrt{K}$ \cite{DeNardis2015,Bouchoule2023}, with $K=(1^{\rm dr}(\pm\theta_F))^2$ the Luttinger parameter of the gas. Notice that there are short-wavelength non-universal corrections to Eq.~\eqref{eq:density-exp} accounting for backscattering processes between the two Fermi edges. Since below we are interested in the evolution of chiral states, we shall not consider these terms in the density expansion.\\

In the presence of an external potential $V(x)$, integrability is lost. A first approximation that is usually considered is the local density approximation (LDA): at each position $x$ the state is a Fermi sea with a $\theta_F(x)$ fixed by the local density $\langle \hat{\rho} (x) \rangle$. We denote filling functions obtained this way as $n_{\rm LDA} (\theta, x)=\Theta(\theta^2-\theta^2_F(x))$. Such filling functions however do not include any quantum fluctuations. Therefore, in the rest of this work, we shall incorporate in the Bethe ansatz description of the gas the fluctuating part \eqref{eq:density-exp} obtained via refermionization. As a result, we obtain an improved initial state, denoted below as $n_\text{Fluct}(x,\theta)$. In doing so, our starting point is Ref.~\cite{Bettelheim2011} (which we revisit in the following section) where the case at strong repulsion $c\to\infty$ has been investigated. 

\section{Tonks-Girardeau limit and Fermi coherent states}\label{sec:tonks}
At strong repulsion $c\to\infty$, the system enters the Tonks-Girardeau regime. For simple observables having no Jordan-Wigner strings, the latter is equivalent to a free Fermi gas, with fermionic degrees of freedom $\theta_i=2\pi I_i/L$ now coinciding with the physical ones. In this limit, as $K\equiv 1$, there is no need to distinguish the fermionic from the bosonic density in eq.~\eqref{eq:density-exp}. Focusing on the right chirality, the density operator is
\be
\delta\hat\rho_+(x)=\hat\psi^\dagger_+\hat\psi_+=\sum_{k>0} e^{\ii k x} \hat{A}^\dagger_{+,k}
\ee
with $\hat{A}_{+,k}$ creating a right-moving phonon, or equivalently a superposition of particle-hole pairs generated around the right Fermi edge $\hat{A}^\dagger_{+,k}=\sum_{p>0} \hat{c}^\dagger_{p} \hat{c}_{p+k}$ \cite{Bouchoule2023}.
\\

Now, as anticipated in the introduction,  Fermi gases in a chiral potential $V(x)$ are found in a coherent state $\ket{V} =\hat{\cal U} | 0 \rangle$, obtained as unitary transformation of the ground state operated by 
\be\label{eq:initialU}
 \hat{\cal U}=\exp\left(\ii \int \Phi(x) \hat{\rho}(x) dx\right).
 \ee

The function $\Phi(x)$ is fixed by the expectation value of density, namely 
\begin{equation}\label{eq:fix-phi}
 \frac{1}{\pi}  \frac{ d\Phi}{dx}=\langle V | \hat{\rho}(x) | V \rangle \equiv\rho_{\rm LDA} (x)= \frac{p_F(x)}{\pi} 
\end{equation}
since no corrections to LDA are expected in the thermodynamic limit for an infinite system. Here, the two Fermi edges $\pm p_F(x)=\pm \sqrt{2(\mu-V(x))}$ and $n_\text{LDA}=\Theta(\theta^2-p_F^2(x))$, such that
\begin{equation}
   \rho_{\rm LDA} (x) =   \int\frac{d\theta}{2 \pi} n_\text{LDA}(\theta,x)      = \frac{p_F(x)}{\pi}.
\end{equation}
$\Phi(x)$ carries also the physical interpretation of a semi-classical phase for long-wavelength modulation of density, similarly to a height field in the bosonization language, see e.g. Refs.~\cite{Giamarchi2003,Cazalilla2004}.\\

Importantly, $\hat{A}_{+,k} \ket{V}=p_k \ket{V}$, with eigenvalues given by the Fourier modes $p_k=\int e^{-\ii kx} d\Phi(x)$. We can therefore conclude that: \emph{i)} $\ket{V}$ is a coherent state for the low-energy quantum fluctuations in~\eqref{eq:density-exp}; \emph{ii)} it has a density of excitations modulated by the external potential $V(x)$ via eq.~\eqref{eq:fix-phi}.\\

At this point, by noticing that the unitary transformation on the chiral fermi operators amounts to an overall phase \cite{Bettelheim2011}
\be
\hat{\cal U} \ \hat\psi_{\pm}(x) \  \hat{\cal U}^\dagger = e^{\mp \ii\Phi(x)} \ \hat\psi_{\pm}(x),
\ee
we can compute two-point function on $\ket{V}$ as
\begin{equation}\label{eq:2-pt-tonks}
    \langle V | \sum_{\sigma=\pm} \hat\psi_\sigma^\dagger (x_1) \hat\psi_\sigma(x_2) | V \rangle  = \frac{2\sin( \int_{x_1}^{x_2} p_F(x') dx' )}{x_1 - x_2}  .
\end{equation}
Taking the Wigner transform of the two-point function, one obtains the occupation function
\begin{equation}\label{eq:nfluctTonks}
    n_{\rm Fluct}(x,\theta) =  \int {dy} e^{\ii y \theta } \ \frac{\sin( \int_{x-y/2}^{x+y/2} p_F(x') dx' )}{\pi y}
\end{equation}
or, equivalently, in terms of the initial potential
\begin{equation}
    n_{\rm Fluct}(x,\theta) =  \int {dy} e^{\ii y \theta } \ \frac{\sin( \int_{x-y/2}^{x+y/2} \sqrt{2(\mu-V(x'))} dx' )}{\pi y},
\end{equation}
where the notation $n_{\rm Fluct}(x,\theta)$ denotes a (fluctuating) filling function taking also negative values, i.e.  carrying the physical interpretation of a quasi-probability of finding a particle in the same spirit of semi-classical Wigner functions \cite{Wigner1932,Hillery1984,Ferrie2011,Moyal1949}. In the limit where the external potential is switched off, then $p_F=\text{const}$ and the integral in eq.~\eqref{eq:nfluctTonks} gives $n_\text{Fluct}=n_\text{LDA}$.  \\

As shown already e.g. in Refs.~\cite{Bettelheim2011,Bettelheim2012,Protopopov2013,protopopov2014,vaness2019,Ruggiero2021,Scopa2023,Dean2019}, the time evolution after a trap release of such fluctuating state can be described by the hydrodynamic equation
\be\label{eq:tonks-hydro}
\partial_t n_\text{Fluct}= - \theta\partial_x n_\text{Fluct},
\ee
with solution $n_\text{Fluct}(t; x,\theta)=n_\text{Fluct}(0; x-\theta t,\theta)$. 

\begin{figure}
\centering
\includegraphics[width=\columnwidth]{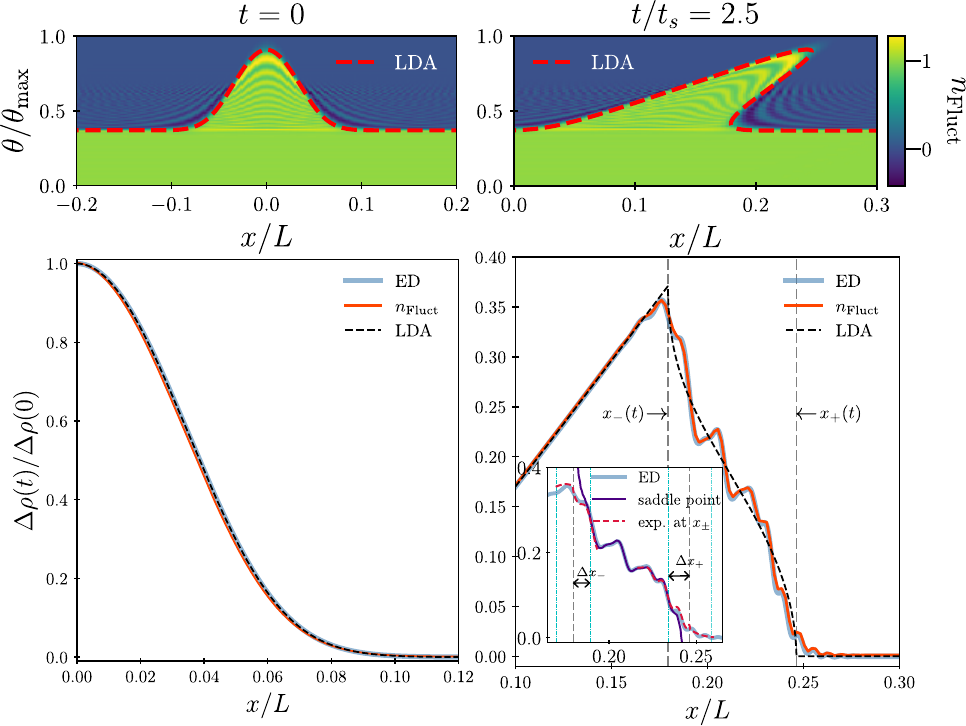}
\caption{
Evolution of a density hump for the Tonks-Girardeau gas. \emph{Top panels}~--~Color plot of $n_{\rm Fluct}$ in phase-space $x$-$\theta$, obtained from eq.~\eqref{eq:nfluctTonks} for times $t=0$ (left) and after the shock time $t>t_s$ (right). Dashed lines show the evolution of $n_\text{LDA}$. \emph{Bottom panels}~--~Corresponding particle density in eq.~\eqref{eq:density-evo} (solid curve), compared to the LDA value (dashed curve) and to exact diagonalization numerics (ED) (thick shaded curve); dashed gray axes mark the boundaries $x_\pm(t)$ of the shock region. \emph{Inset:}~ED is compared to the saddle point approximation in eq.~\eqref{eq:qghd_ripples} (dark solid curve) and to the expansions in eq.~\eqref{eq:caustic} near the shock points (dashed curves); dash-dotted cyan axes mark the regions $\Delta x_\pm$ around the shock points. In this figure, the density hump is generated by $\mu-V(x)=0.178+0.8\exp(-x^2/20^2)$, and numerics are done for a finite-size system of $L=4000$ sites. Density is rescaled as $\Delta \rho(t)/\Delta\rho(0)= \frac{\langle\hat\rho(t,x)\rangle-\rho_0}{\langle\hat\rho(0,0)\rangle-\rho_0}$, where $\rho_0=\sqrt{2\mu}/\pi$ is the LDA background value.
}\label{fig:tonks-example}
\end{figure}
\subsection{Corrections to LDA density during the gas expansion}
Our focus is on the profile of particle density during the gas expansion of Fig.~\ref{fig:shock}, obtained at each time as
\be\label{eq:density-evo}
\langle\hat\rho(t,x)\rangle=\int \frac{d\theta}{2\pi} n_\text{Fluct}(x-\theta t,\theta),
\ee
and particularly to determine the deviations from its LDA value $\rho_\text{LDA}(t,x)=\int \frac{d\theta}{2\pi} n_\text{LDA}(x-\theta t,\theta)$. In the Tonks-Girardeau limit, such deviations can be analytically determined. Below, we report the main passages and comment on the results, leaving technical derivations and further details to Appendix~\ref{app:correction-lda-tg}.\\

\begin{figure*}[t!]
    \centering
    (a) \hspace{3.5cm} (b)\\
    \includegraphics[width=0.9\textwidth]{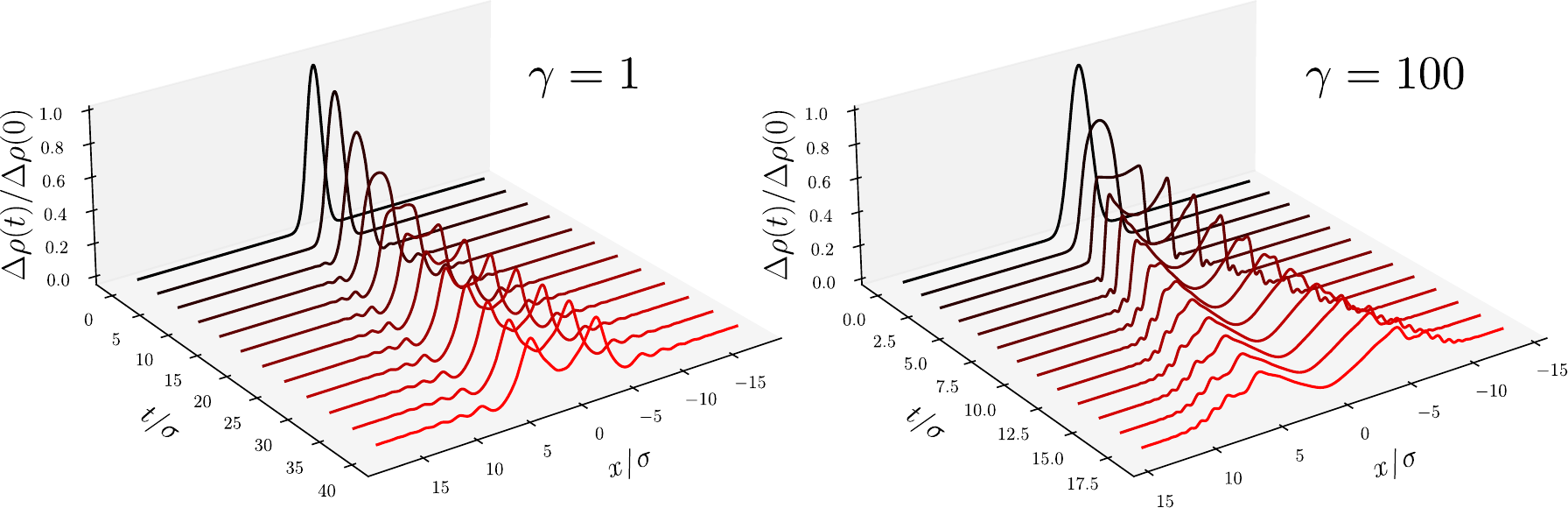}
    \caption{Evolution of the density hump $\Delta\rho(t)/\Delta\rho(0)=\frac{\langle\hat\rho(t,x)\rangle -\rho_0}{\langle\hat\rho(0,0)\rangle-\rho_0}$ for the Lieb-Liniger gas at $\gamma =1$ (\emph{left panel}) and $\gamma = 100$ (\emph{right panel}), $\gamma = c/\rho_0$ where $\rho_0$ is the LDA background density. Post shock, the density exhibits ripples of different and distinct characteristics in the two regimes. For low values of $\gamma$ (\emph{left}), low-frequency ripples are observed near the propagation fronts, which are identified with semi-classical waves, whereas in the Tonks regime (\emph{right}) the ripples are dominated by high-frequency oscillations spreading all over the shock region, which characterize what we refer to as quantum ripples. Both density humps were generated by the potential $V(x) - \mu = \alpha + \beta \exp(-x^2/\sigma^2)$, with $\alpha,\beta$ chosen such that $0.05 \leq \rho(t=0,x) \leq 0.2$ and $\sigma =25, 20$ for the left and right plots, respectively. }
    \label{fig:isometric_density}
\end{figure*}


 To begin with, we associate a vector $\vec{x}= (x,\theta)$ to each phase-space point, and define the Fermi contour as the geometric locus satisfying at $t=0$
\be
\Gamma_0=\{(x,\theta) : \ \theta^2=p_F^2(x)\}.
\ee
$\Gamma_0$ can be described as  a simple curve in phase space since the system is characterized by two symmetric Fermi edges $\pm p_F(x)$ at each $x$. In order to take advantage of this, we introduce a coordinate $s(\vec{x})\in \mathbb{R}$ along  $\Gamma_0$ such that (with slight abuse of notation) 
\be
\Gamma_0(s)=\{ s(\vec{x})\ :\ \theta_0(s)=p_F(x_0(s))\}.
\ee
For simplicity, below we specialise to the right-moving front of density $x>0$ (the left-moving front is obtained by symmetry). It is then possible to choose the coordinate $s$ to be
\begin{equation}\label{eq:s-coo}
    s(x_0) = \int^{x_0}_{-\infty} \frac{d x'}{p_F(x')}, 
\end{equation}
which conveniently replaces the spatial coordinate $dx$ with the propagating time $ds=dx/p_F(x)$ on the non-flat Fermi surface, see e.g. Refs.~\cite{Brun2017,Brun2018,Scopa2020}. This curve holds also another physical meaning, being directly related to a phase-space quantization condition, see Refs.~\cite{berry_semi-classical_1997,arnold1978} and Appendix~\ref{app:correction-lda-tg}.  In terms of this coordinate, the dynamics generated by eq.~\eqref{eq:tonks-hydro} is encoded in the equation of motion: ($\theta_t(s)\equiv \theta_0(s)$ is conserved)
\be\label{eq:eq-of-motion}
x_t(s)=x_0(s) +t \theta_0(s)
\ee
which can be solved requiring that $x_t(s)\equiv x$, yielding the replacement of $x_0$ with $x-\theta t$ in eq.~\eqref{eq:s-coo}.\\

 We then write the time-evolved density in eq.~\eqref{eq:density-evo} as
\begin{align}\label{eq:density-fluc0}
     \langle \hat\rho(t,x) \rangle &= \iint \frac{d\theta \, dy }{2\pi^2} \sqrt{\left|\frac{ds(x_1)}{dx}\right|\left|\frac{ds(x_2)}{dx}\right|}\nn \\
     &\quad \times \frac{ \sin\left(\int_{x_1}^{x_2} dx_0 \ \frac{dx'_t}{dx_0} p_F(x'_t) \right) e^{\ii \theta y}}{(s(x_2) - s(x_1))} \, ,    
\end{align}
with backward-evolved coordinates 
\be\label{x12}
x_{1,2}=x - \theta t \mp y/2,
\ee
and time-dependent phase difference
\begin{align}\label{eq:time-dep-WKB}
\Delta\Phi_t(x_1,x_2) &=\int_{x_1}^{x_2} dx_0(s) \ \frac{dx'_t(s)}{dx_0(s)}\; p_F(x'_t(s))\nn \\
&=\int_{x_1}^{x_2} dx'\ p_F(x') + \frac{t\,  p_F^2(x')}{2}\bigg\vert_{x_1}^{x_2}.
\end{align}
The second term in \eqref{eq:time-dep-WKB} is nothing but the dynamical phase cumulated by a single particle with momentum $p_F(x)$ that freely expands for a time $t$ \cite{Dubail2017,Ruggiero2019,Scopa2023}. The jacobians are easily evaluated as (see Appendix~\ref{app:correction-lda-tg})
\be\label{eq:jacobians}
\frac{ds(x_{1,2})}{dx}=\frac{1}{p_F(x_{1,2})[1 + t p_F'(x_{1,2}) ]}.
\ee
One can then introduce the semi-classical action
\be\label{eq:semiclass-action}
S(\theta,y)=-\theta y + \Delta\Phi_t(x_1,x_2)
\ee 
and look for the saddle point configuration $\nabla S(\theta,y)=0$. Away from singularities, the saddle point fixes the major contribution to eq.~\eqref{eq:density-fluc0} as coming from $y\to 0$,
\begin{align}\label{eq:density-fluc1}
     \langle \hat\rho(t,x) \rangle \approx &\lim_{y\to 0} \sum_{\substack{x_1\in \{x^\star+y/2\}\\ x_2\in \{x^\star-y/2\}}}\sqrt{\left|\frac{d s(x_1)}{dx}\right|\left|\frac{d s(x_2)}{dx}\right|}\nn \\  &
    \qquad\times\frac{ \sin\left(\Delta\Phi_t(x_1,x_2) \right)}{2\pi(s(x_2)- s(x_1))}
\end{align}
 with $x_{1,2}=x^\star \pm y/2$ projected to the Fermi contour value, that is
\be\label{xstar}
x^\star= x- t p_F(x^\star).
\ee
Clearly, $x^\star\equiv x^\star(t;x)$. This self-consistent equation can display multiple solutions. Precisely, multiple solutions are found in the \emph{shock region} $x_-(t)\leq x\leq x_+(t)$ with boundaries $x_\pm(t)$ satisfying
\be\label{eq:shock-points}
p'_F[x^\star(t;x)]\bigg\vert_{x=x_\pm(t)}=\infty,
\ee
as also explained in Appendix~\ref{app:correction-lda-tg}.\\

It is then easy to see that for $x \notin  [x_-(t);x_+(t)]$, the limit $y\to 0$ corresponds to $x_1\to x_2$, and one recovers the LDA result for the density from eq.~\eqref{eq:density-fluc1} since $\lim_{x_1\to x_2} \Delta\Phi_t(x_1,x_2)\approx p_F(x^\star) y$. On the other hand, inside the shock region corrections to LDA density are observed, as derived below. For convenience, we separate the $x\approx x_\pm(t)$ region from the rest of the shock region, $x_-(t)<x<x_+(t)$, discussion.
\begin{figure*}
\begin{minipage}{0.25\textwidth}
\textbf{(b)}
\end{minipage}\begin{minipage}{0.75\textwidth}
\textbf{(a)}
\end{minipage}\\
\includegraphics[width=\textwidth]{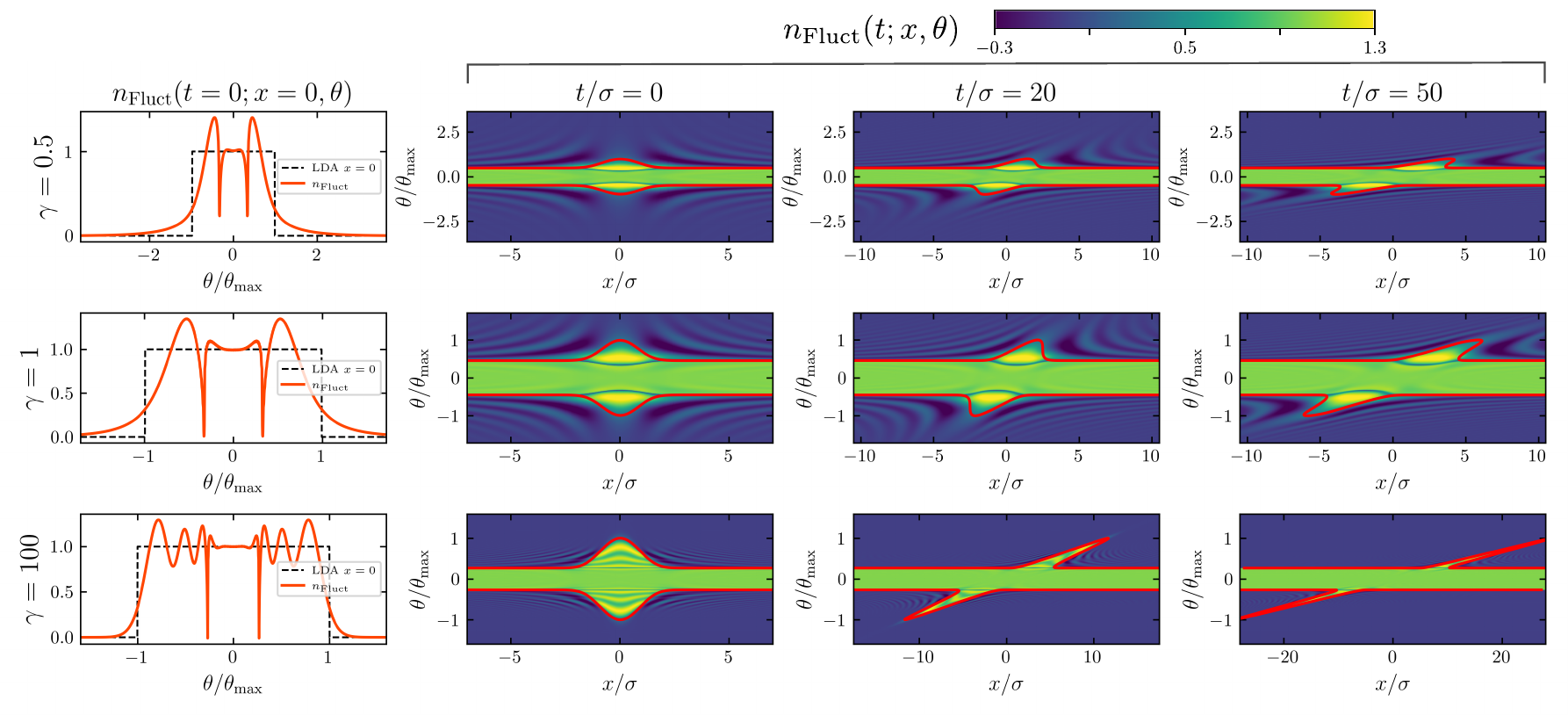}
  \caption{\textbf{(a)}~--~Color map of $n_{Fluct}$ in the $x$--$\theta$ plane for different $\gamma=0.5,1,100$ (rows) and different times $t/\sigma=0,20,50$ (columns). Red solid lines show the evolution of $n_\text{LDA}$ obtained from zero-entropy GHD \cite{Doyon2017}. \textbf{(b)}~--~Cross-section of $n_{\rm Fluct}$ at $t=0$ and $x=0$, shown as function of $\theta$. In this figure, $\theta_{\rm max}=\max(\theta_F(t=0,x))$ and the density hump is initially generated by the potential $V(x)-\mu=\alpha +\beta \exp(-x^2/\sigma^2)$ with $\alpha,\beta$ chosen such that $0.05\leq \rho(t=0,x)\leq 0.2$ and $\sigma=20$.
  }\label{fig:nfluct-gammas}
\end{figure*}
\subsubsection{Saddle point contribution in the shock region}
In the region $x_-(t)<x<x_+(t)$, eq.~\eqref{xstar} has three different solutions $x^\star_j$ satisfying $x^\star_1<x^\star_2<x^\star_3$. They correspond to 12 terms in the sum of eq.~\eqref{eq:density-fluc1}. The ``diagonal'' contributions $x_{1,2}=x^\star_j\pm y/2$ reproduce the LDA result, that is

\be\label{eq:LDA-dens-tonks}
\rho_\text{LDA}(t,x)=\frac{p_F(x^\star_1)-p_F(x^\star_2)}{2\pi} +\frac{p_F(x^\star_3)}{2\pi} + \frac{\rho_0}{2},
\ee
$\rho_0=\sqrt{2\mu}/\pi$, while the non-diagonal terms give rise to LDA corrections
\begin{align}
    \langle \delta\hat\rho(t,x)\rangle\bigg|_{x^\star_i,x^\star_j} &\approx  \sqrt{\left|\frac{d s(x^\star_i)}{dx}\right|\left|\frac{d s(x^\star_j)}{dx}\right|}\nn \\
   & \quad\times\frac{ \sin\left( \Delta\Phi_t(x^\star_i,x^\star_j)  - {\cal I}_{i j} \frac{\pi}{2}\right)}{2\pi(s(x^\star_i )-s(x^\star_j ))} \, , \nn\\
\end{align}
where ${\cal I}_{i j}$ is the Maslov index along the path $s(x^\star_i)\to s(x^\star_j)$ on the Fermi surface, see Appendix~\ref{app:correction-lda-tg} for more details. Combining all terms together, we obtain

\begin{widetext}
\begin{eqnarray}
\label{eq:qghd_ripples}
       \langle \delta\hat\rho(t,x>0) \rangle 
      &=& - J_{12}\frac{ \cos\left(\Delta\Phi_t(x^\star_1,x^\star_2)\right)}{\pi |s(x^\star_1) -s(x^\star_2) |} +J_{23}\frac{ \cos\left( \Delta\Phi_t(x^\star_2,x^\star_3) \right)}{\pi |s(x^\star_2) - s(x^\star_3) |}  - J_{13}\frac{ \sin\left( \Delta\Phi_t(x^\star_1,x^\star_3) \right)}{\pi |s(x^\star_1) - s(x^\star_3) |},
\end{eqnarray}
\end{widetext}
where the shorthand $J_{ij} =\sqrt{|ds(x^\star_j)/dx| |ds(x^\star_i)/dx|}$ has been used for convenience. Notice that the same analytical expression in eq.~\eqref{eq:qghd_ripples} can be obtained from quantum GHD~\cite{Scopa-Dubail-tonksripples, Ruggiero2021}. It is easy to see that $\Delta\Phi_t( x^\star_i,x^\star_j)$ is nothing but the number of particles in the hump between points $s(x^\star_i)$ and  $s(x^\star_j)$ along the time-evolved Fermi surface (cf. Fig.~\ref{fig:fig2}), see Appendix~\ref{app:correction-lda-tg} for more details. Eq.~\eqref{eq:qghd_ripples} gives a non-vanishing correction only if $\Delta\Phi_t\sim {\cal O}(\Delta N)$, $\Delta N$ the total number of particles in the hump, therefore it can be understood as a quantum correction to LDA density according to the classification of Sec.~\ref{sec:intro}.

\subsubsection{Expansion at the shock points}
Close to the shock points $x\approx x_\pm(t)$, the limit $y\to 0$ becomes more complicated. If we approach the shock points from inside the shock region, namely $x= x_\pm(t)\mp\epsilon$ for $\epsilon\to 0$,  two roots in eq.~\eqref{xstar} coincide and lead to contact singularities in eq.~\eqref{eq:qghd_ripples}. On the other hand, if $x\to x_\pm \pm\epsilon$ then eq.~\eqref{xstar} is single valued and \eqref{eq:density-fluc1} reproduces the LDA density.\\
As noted already in Refs.~\cite{Bettelheim2011,Bettelheim2012,Doyon2017,gouraud2024quantum}, a good characterization of this region requires an expansion of the phase difference in eq.~\eqref{eq:density-fluc0} beyond saddle-point contribution, that is
\be
\Phi(x_2)-\Phi(x_1)\approx p_F(x^\star) y + \frac{p_F''(x^\star) y^3}{24}.
\ee
Using this, one finds universal corrections at the edges of the shock region in the form of a Airy kernel \cite{Doyon2017,Fagotti2017,Bettelheim2011,Bettelheim2012,gouraud2024quantum}
\begin{eqnarray}
    \langle \hat\rho(0,x) \rangle
 &=& \int \frac{d\theta}{2\pi \ii} \text{Ai}_1\left( \frac{\theta- p_F(x^\star)}{\sqrt[3]{\ii p''_F(x^\star)/8}} \right),
\end{eqnarray}
$\text{Ai}_1(\cdot)$ being the incomplete Airy function defined as $\text{Ai}_1 (x) = \int d q/(2\pi \ii) \exp(\ii q x +  \ii q^3/3)/q$. By recasting the expression in terms of Fermi positions at $t=0$, an  equivalent expression close to $x_\pm(t)$ and valid for any time $t$ after the shock formation time is found~\cite{Bettelheim2012}
 \begin{eqnarray}
    \langle \hat\rho(t,x) \rangle 
&=& \int \frac{d\theta}{2\pi} \text{Ai}_1\left(  \frac{x- x^\star}{\sqrt[3]{x''_F(\theta^\star)/(8\ii )}} \right)
\end{eqnarray}
this expression is single-valued for $x\approx x_\pm (t)$. Here, $x_F(\theta)$ is defined such that $p_F(x_F(\theta))=\theta$, and $x_F(\theta^\star)\equiv x^\star=x-\theta^\star t$. By expanding this result for small $\delta x_\pm =\mp (x - x_\pm(t))$, one obtains the following correction to the LDA density
\begin{eqnarray}\label{eq:caustic}
     \langle \delta\hat\rho(t,x\approx x_\pm(t)) \rangle &=& (\kappa^\pm)^2 \bigg[  \delta x_\pm [\text{Ai}(\kappa^\pm \delta x_\pm)]^2  \nn \\
     &&\quad\qquad -\frac {  [\text{Ai}'(\kappa^\pm \delta x_\pm)]^2} {\kappa^\pm}\bigg]
\end{eqnarray}
where $\kappa^\pm = \left[ 2 \ii / (x_F''(\theta^\star) )\right]^{1/3}\sim {\cal O}(\ell^{2/3})$. This result can be combined with the LDA solution \eqref{eq:LDA-dens-tonks} and with regular terms of~\eqref{eq:qghd_ripples} to correctly determine the behavior of density at the shock points. Finally, we briefly comment on the range of validity of eq.~\eqref{eq:caustic}. Since these corrections are centered around the shock points $x_\pm(t)$ where the local number of excess particles becomes very small, $\lim_{y\to 0}\Delta\Phi_t= 0$, they cannot belong to the quantum regime. On the other hand, the total number of particles contained in the density hump $\Delta N$, provides a useful estimate as to the width of the Fermi surf and sets the range of validity to be approximately $|\delta x_\pm| \leq \Delta x_{\pm} \sim (\Delta N)^{2/3}/ \kappa^\pm$ \cite{Bettelheim2012}, which is indeed finite in the semi-classical regime. 

In Fig.~\ref{fig:tonks-example}, we show the evolution of the fluctuating initial state in eq.~\eqref{eq:tonks-hydro} and the corresponding evolution of the particle density \eqref{eq:density-evo} for the setup illustrated in Fig.~\ref{fig:shock}. As one can see, the density curve obtained through $n_\text{Fluct}$ shows oscillations on top of the LDA value inside the shock region, and it perfectly matches the exact diagonalization numerical data. In the inset of Fig.~\ref{fig:tonks-example}, the analytical results for the density ripples in eqs.~\eqref{eq:qghd_ripples} and \eqref{eq:caustic} are compared with exact numerics. We see that around the shock points $x_\pm(t)$ one finds the regions $\Delta x_\pm$ inside which eq.~\eqref{eq:caustic} gives a good approximation of \eqref{eq:density-evo} while \eqref{eq:qghd_ripples} shows contact singularities. Outside these regions, the saddle point approximation \eqref{eq:qghd_ripples} is regular and it reproduces the numerical data extremely well.


\begin{figure*}
\includegraphics[width=\textwidth]{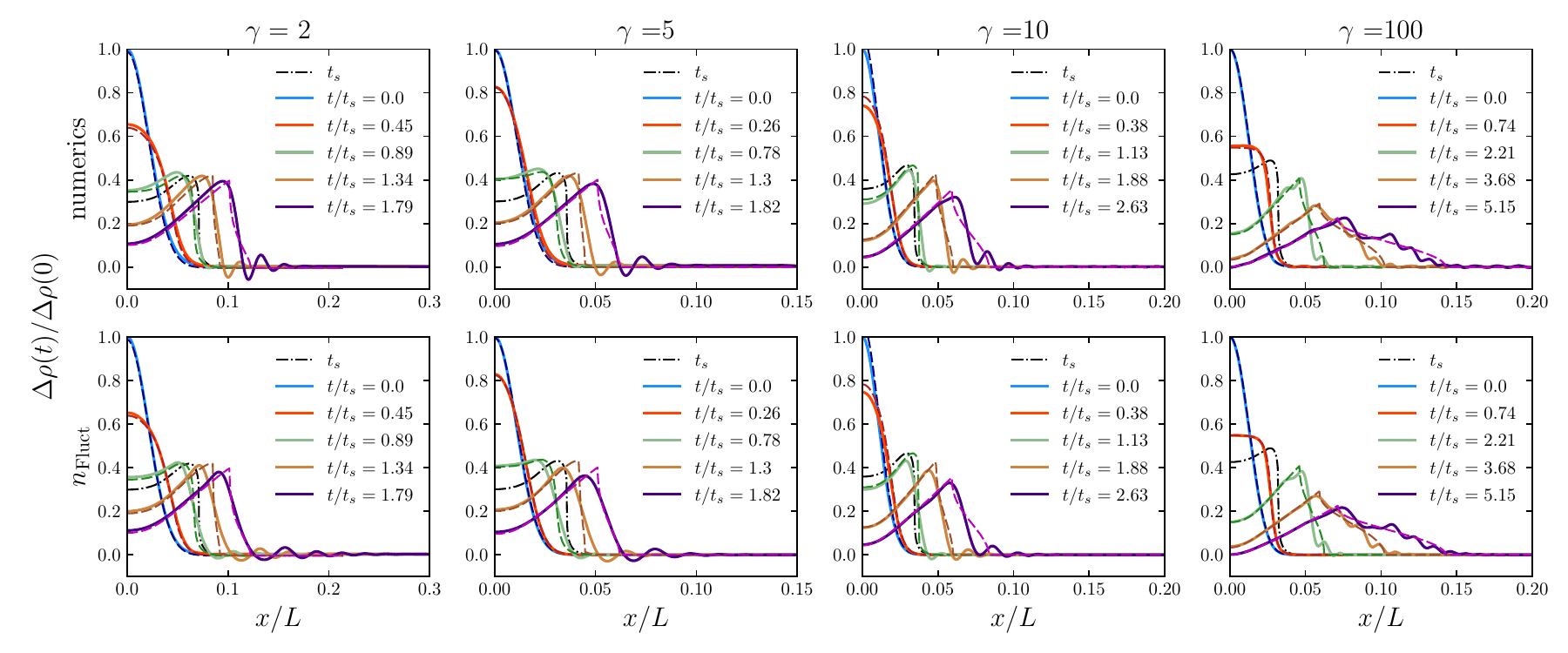}
  \caption{
Evolution of the density hump $\Delta\rho(t)/\Delta\rho(0)=\frac{\langle\hat\rho(t,x)\rangle -\rho_0}{\langle\hat\rho(0,0)\rangle-\rho_0}$ for the Lieb-Liniger gas at different $\gamma=c/\rho_0$, with $\rho_0$ the LDA background density. $\gamma$ increases from leftmost to rightmost panels. In each panel we find: \emph{i)}~dashed-line curves showing the LDA value of density, obtained with zero-entropy GHD; \emph{ii)}~dot-dashed curve showing the density profile at the shock time $t_s$, estimated as tangent point in the Fermi contour evolution (see Ref.~\cite{Doyon2017}). \emph{iii)}~fluctuating density: \emph{top row}~--~numerical data obtained with tDMRG \cite{itensor}, see main text for a discussion; \emph{bottom row}~--~prediction obtained using $n_{\rm Fluct}$ and Euler GHD, see eq.~\eqref{eq:density-evo}; we checked that higher-order corrections to Euler GHD can be neglected. In this figure, the density hump at $t=0$ is generated by the potential $V(x)-\mu=\alpha+\beta \exp(-x^2/\sigma^2)$ with $\alpha,\beta$ chosen such that $\rho(0,0)=0.2$ and $\rho_0=0.05$; $\sigma=20$ for $\gamma=5,10,100$, $\sigma=40$ for $\gamma=2$. The lattice model used for the numerics has $L=1000$ sites and open boundaries conditions. 
  }\label{fig:density-gammas-DMRG}
\end{figure*}

\section{Fluctuating initial state at finite interaction}\label{sec:interact-method}
At finite interactions, a (universal) fermionized description of the quantum fluctuations of the Bose gas is given by Eq.~\eqref{eq:density-exp}. For non-homogeneous systems,
\be\label{eq:densB-densF}
\hat\rho(x)=\sqrt{K(x)} \hat\rho_F(x)
\ee
where we denoted $\hat\rho_F=\sum_\sigma\hat\psi^\dagger_\sigma\hat\psi_\sigma$ and $K(x)$ is the LDA value of the Luttinger liquid constant, related to the local compressibility of the gas \cite{Giamarchi2003,Cazalilla2011}. It is then possible to build a coherent Fermi state as in Sec.~\ref{sec:tonks} but now for the refermionized degrees of freedom, see also Ref.~\cite{protopopov2014}. We write this state as 
\begin{equation}
    | V \rangle= \exp \left( \ii \int dx   \ \Phi_{\rm int}(x) \hat{\rho}_F(x) \right) | 0 \rangle
\end{equation}
with function $\Phi_{\rm int}(x)$ again fixed by the expectation value of density. Using eq.~\eqref{eq:densB-densF}, we have  
\begin{equation}\label{eq:PhiInt}
  \frac{ \sqrt{K(x)}}{\pi} \frac{d \Phi_{\rm int}}{dx} =\langle V | \hat{\rho}(x) | V \rangle = \rho_{\rm LDA} (x)  
\end{equation}
implying that $d \Phi_{\rm int}/dx = \pi\rho_\text{LDA}(x)/\sqrt{K(x)}$ and therefore, repeating the previous steps 
\begin{equation}\label{eq:nFluct-LL-0}
    n_{\rm Fluct}(x,k) =  \int dy e^{\ii y k}  \ \frac{\sin\left(\int_{x-y/2}^{x+y/2}  \frac{\pi \rho_\text{LDA}(x')}{\sqrt{K(x')} } dx' \right)}{\pi y}.
\end{equation}
This expression is universal, as it depends only on the value of density $\rho_\text{LDA}$ and on the Luttinger parameter $K$ of the one-dimensional gas. Specifying to the Lieb-Liniger model, we can express the phase in eq.~\eqref{eq:nFluct-LL-0} as $d\Phi_\text{int}/dx= p_F^{\rm Dr}/\pi$. Here we have introduced the physical momentum, which differs from the momentum $p_F$ of a noninteracting system due to scattering shift in eq.~\eqref{eq:BetheEq}. The physical dressing operation (notice that this ``Dr'' operation is different from ``dr'' appearing in eq.~\eqref{eq:dr-operation}) of a function $h(\theta)$ respect to any filling $n$, is written as 
\begin{eqnarray}\label{eq:pDress}
    h_{[n]}^{\rm Dr}(\theta) = h(\theta) + \int  \frac{d \alpha}{2 \pi} \ \phi(\theta - \alpha)  (h')^{\rm dr}_{[n]}(\alpha) n(\alpha) .
\end{eqnarray}
We denote with $p^{\rm Dr}_F(x)\equiv p^{\rm Dr}_{[n]}(\theta_F(x))$ the dressed Fermi momentum computed at the Fermi point, which is related to the LDA density as
\begin{equation}
  \rho_{\rm LDA} (x) =\int^{\theta_F(x)}_{-\theta_F(x)} \frac{d\theta}{2\pi} 1^{\rm dr}_{[n_\text{LDA}]}(x,\theta)=    \int_{-p^{\rm Dr}_F(x)}^{p^{\rm Dr}_F(x)} \frac{dp}{2\pi} .
\end{equation}
Using the relation
\be
p^{\rm Dr}_F(x)=\theta_F^{\rm dr}(x)\sqrt{K(x)},
\ee
with $\theta^{\rm dr}_F(x)\equiv\theta_{[n]}^{\rm dr}(x,\theta_F(x))$, and dressing in eq.~\eqref{eq:dr-operation} performed using $n_\text{LDA}(x,\theta)$ at each position $x$, we rewrite \eqref{eq:nFluct-LL-0} as
\begin{equation}\label{eq:nFluct-LL}
    n_{\rm Fluct}(x,k) =  \int dy e^{\ii y k}  \ \frac{\sin\left(\int_{x-y/2}^{x+y/2} dx' \theta^{\rm dr}_F(x') \right)}{\pi y}.
\end{equation}
This object carries the physical interpretation of quasi-probability in the same sense of eq.~\eqref{eq:nfluctTonks}, but for refermionized degrees of freedom rather than for the physical ones. It becomes clear then that the fluctuating macrostate in \eqref{eq:nFluct-LL} is related to the macrostate of rapidities $ n_{\rm Fluct}(x,\theta)$ via a (nonlinear) reparametrization of the quasimomentum space
\be\label{eq:eqmapping}
n_{\rm Fluct}(x,\theta)=n_{\rm Fluct}(x,k\equiv p_{[n_{\rm Fluct}]}^{\rm Dr}(\theta,x)/\sqrt{K(x)}).
\ee
This also shows that the density (and all other thermodynamic quantities) can be obtained as 
\begin{align}
    \rho(x)  & = \sqrt{K(x)} \int \frac{dk}{2 \pi} n_{\rm Fluct}(x,k)  \nonumber \\& =  \int  \frac{d\theta}{2 \pi} 1_{[n_{\rm Fluct}]}^{\rm dr}(x,\theta)\ n_{\rm Fluct}(x,\theta) .
\end{align}

Eq.~\eqref{eq:nFluct-LL} together with the mapping to standard rapidities, eq. \eqref{eq:eqmapping}, (or equivalently eq.~\eqref{eq:nFluct-LL-0} for the non-integrable gas) represent our main result. It specifies the quantum fluctuating initial state for the interacting Bose gas in terms of universal quantities. Its time evolution requires the knowledge of a Boltzmann-like equation in phase space \cite{panfil2023thermalization,bertini2015pretherm,durnin2021nonequilibrium,piqueres2023integrability,essler2014quench,delvecchiodv2022,bastianello2021hydrodynamics}. For instance, one can consider the  kinetic equation for the refermionized quasiparticles~\eqref{eq:nFluct-LL-0} proposed in Ref.~\cite{protopopov2014}. For the Lieb-Liniger gas, it can be easily carried out using GHD, as discussed in the following paragraph.

\subsection{Hydrodynamic evolution}\label{sec:hydro-evo}
In this paragraph, we discuss the evolution of the setting in Fig.~\ref{fig:shock} for the interacting Lieb-Liniger gas $\gamma=c/\rho_0\sim{\cal O}(1)$, with $\rho_0$ the LDA background density. The latter is evolved using standard GHD equations \cite{Castro-Alvaredo2016,Bertini2016}
\be\label{eq:EulerGHD}
\de_t n_{\rm Fluct}(t;x,\theta) + v^{\rm eff}_{[n_{\rm Fluct}]}(\theta) \de_x n_{\rm Fluct}(t; x,\theta ) =0 ,
\ee
with effective velocity $v^{\rm eff}_{[n]}(\theta)= \theta_{[n]}^{\rm dr}(\theta)/1_{[n]}^{\rm dr}(\theta)$. Higher-order corrections to eq.~\eqref{eq:EulerGHD}, in particular dispersive terms $\propto\de_x^3$, are present, see eq. \eqref{eq:EulerGHD2}. Their expression has been determined in Ref.~\cite{denardis_doyon_dispersive}. 
However, we checked that these terms remain small in the setup of Fig.~\ref{fig:shock} for the times that we investigate, and thus will be neglected in the following. Moreover we remark that for small initial modulation of the density, namely when we can approximate the Luttinger parameter to be space independent $K(x) \simeq K_0$, we can also write the evolution directly in the coordinate $k$ as
\begin{equation}
    \partial_t n(t;k,x) +  \sqrt{K_0} \ \partial_k \theta^{\rm Dr}_{[n_{\rm Fluct}]}   \partial_x n(t;k,x) = 0 .
\end{equation}

We recall that applying Euler GHD \eqref{eq:EulerGHD} to $n_{\rm LDA}(t; x,\theta)$ yields to the so-called zero-entropy (or zero-temperature) GHD \cite{Doyon2017}, with Fermi points being the only hydrodynamic modes governing the time evolution. Indeed, by applying spatial and time derivatives to the LDA state give Dirac delta functions $\delta(\theta- \theta_F^\zeta(t,x))$  around the Fermi points $\theta_F^\zeta(t,x)$, allowing to recast $(\de_t+v^{\rm eff}\partial_x) n_{\rm LDA}=0$ as
\begin{align}\label{eq:zeroEGHD}
 \sum_{\zeta = \pm}  \zeta \delta (\theta - \theta_F^\zeta )  [ \partial_t \theta_F^{\zeta} + v^{\rm eff}(\theta_F^\zeta)  \partial_x \theta_F^{\zeta} ]=0.
\end{align}
Here, we assumed that there is a single Fermi sea with boundaries $\theta_F^\pm(t,x)$, the generalization to multiple Fermi seas proceeds similarly, see Ref.~\cite{Doyon2017}.
However, this is not the case when considering $n_{\rm Fluct}(t; x,\theta)$ as initial state, and retaining the full $\theta$-dependence becomes necessary to properly characterize the hydrodynamic evolution. \\

In order to illustrate this, we focus on the simplified case where the spatial inhomogeneity is large $\ell\sim\de_x\rho_\text{LDA}/\rho_\text{LDA}\gg 1$, so that only the semi-classical ripples of $n_{\rm Fluct}$ need be considered. In this case, $\partial_x \theta_F \sim {\cal O}(1/\ell)$ and thus one can expand the phase difference entering in eq.~\eqref{eq:nFluct-LL} as
\begin{eqnarray}\label{eq:phaseExpansion}
    \int^{x+y/2}_{x-y/2} \dr \theta_F(x') dx' \approx \theta^{\rm dr}_F(x) y + \frac{y^3}{24} \partial_x^2 \dr \theta_F(x).
\end{eqnarray}
This expansion motivates the definition of $\alpha \equiv \partial_x^2 \theta^{\rm dr}_F / 8$, and can be reinserted into $n_{\rm Fluct}$ \eqref{eq:nFluct-LL} to determine
\begin{eqnarray}\label{eq:nfluctAiry}
    n_{\rm Fluct} &=& \int \frac{dy}{ \pi y} e^{\ii  y k}  \sin \left(    \dr \theta_F y + \frac{y^3}{24} \partial_x^2 \theta^{\rm dr}_F  + \ldots \right)   \nn\\
    &=& \int dq \left[\Theta(k + \dr \theta_F - q \alpha^{\frac 1 3}) - \Theta( \dr \theta_F - k + q \alpha^{\frac 1 3} ) \right] \nn\\
    &&\times \Ai(q).
\end{eqnarray}
Now, it follows that the application of a spatial derivative acts only on $\dr \theta_F$ and $\alpha$ so that
\begin{eqnarray}\label{eq:expansiondelta}
    \partial_x \Theta(k + \dr \theta_F - q \alpha^{\frac 1 3}) &=& \delta(k + \dr \theta_F - q \alpha^{\frac 1 3}) \nn\\
    &&\times(\partial_x \dr \theta_F - q \frac{ \partial_x^3 \dr \theta_F}{24 \alpha^{\frac2 3}}).
\end{eqnarray}
As a result, it follows that, for the case of  generic number of Fermi points, labelled by the index $\zeta$, and with signs $s_\zeta$, such that Fermi point position is  $\theta_F^\zeta(t,x)$, we obtain 
\begin{equation}
   \partial_x n(k= \theta^{\rm dr}) = \sum_{\zeta}  \ s_\zeta 
\frac{\delta \theta^{\rm dr} (\theta_F^\zeta)}{\delta (\theta_F^\zeta)}  \frac{\Ai(\frac{k +  \theta^{\rm dr}(\theta_F^\zeta) }{\alpha^{1/3}} ) }{\alpha^{1/3}} \partial_x \theta_F^\zeta   + {\cal O}(\partial_x^3),
\end{equation}
and, since $  \int d k \alpha^{-1/3} \Ai(k \alpha^{-1/3}) = 1$,  we find
\begin{equation}
    \frac{\Ai(\frac{k +  \theta^{\rm dr}(\theta_F^\zeta) }{\alpha^{1/3}} ) }{\alpha^{1/3}}  = \delta(k + \theta^{\rm dr}(\theta_F^\zeta) ) + {\cal O}(\ell^{-2/3}).
\end{equation}
Hence, at small $\alpha \sim 1/\ell^2$ (or equivalently for large spatial variations $\ell\gg 1$), zero-temperature Euler GHD \eqref{eq:zeroEGHD} is indeed recovered. 
However, higher-order corrections coming from the second term in eq.~\eqref{eq:expansiondelta} generates also \emph{higher-order hydrodynamics terms}, precisely dispersive terms at leading order, correcting the Euler evolution of the Fermi points
\begin{eqnarray}\label{eq:dispersive-fermi-points}
   \sum_{\zeta}  \frac{ s_\zeta}{24 \alpha} \Ai''\left( \frac{k + \dr\theta_F(\theta_F^\zeta)}{\alpha^{\frac 1 3}}\right) \partial_x^3 \dr\theta_F(\theta_F^\zeta).
\end{eqnarray}
Here, we used the known identity $x \Ai(x) = \Ai''(x)$. Importantly, the hydrodynamic coefficient in eq.~\eqref{eq:dispersive-fermi-points} has a finite limit in terms of derivative of Dirac delta for large $\ell$, that is 
\begin{eqnarray}
        \frac{\Ai''(\frac{k +  \theta^{\rm dr}(\theta_F^\sigma) }{\alpha^{1/3}} ) }{\alpha} \overset{\ell\gg 1}{\approx} \delta''(k + \theta^{\rm dr}(\theta_F^\sigma) ).
\end{eqnarray}
Remarkably, this shows that the Euler evolution of $n_{\rm Fluct}$ contains dispersive hydrodynamics terms even at large $\ell \gg 1$, once expressed in terms of only LDA quantities, such as the Fermi points $\theta^\zeta$. It moreover shows that even for large $\ell$ (or small $c$, see Sec.~\ref{section:semi-classical_limit}), the phase terms of $n_{\rm Fluct}$ cannot be neglected, if one wants to go to times which are beyond Euler scale.

\subsection{Gross-Pitaevskii limit }\label{section:semi-classical_limit}
An important question concerns what happens to $n_{\rm Fluct}$ in the limit of weak interaction $c\to 0^+$, especially at large density $\rho \sim 1/c$.
In this limit, it is known that the ground state of \eqref{eq:lieb-liniger-H} is a quasicondensate \cite{popov_theory_1977,prolhac_ground_2017}, and it is described by a bosonic mean-field single-particle wavefunction $\Psi(x,t)$ playing the role of order parameter for the superfluid regime. Its dynamics is captured by the Gross-Pitaevskii equation (GPE) \cite{damski2004,hoefer2006,meppelink2009,Peotta2014,10.21468/SciPostPhysCore.6.3.064,del_Vecchio_del_Vecchio_2020,Koch2022}, which is purely classical and reads as
\begin{equation}\label{eq:gross_pita}
    \ii\partial_t \Psi = \left(-\frac{\de_x^2}{2} + V(x)  -  \mu +  c|\Psi|^2\right)\Psi.
\end{equation}
In the homogeneous case $V(x) =0$, GPE is known in the mathematical literature as Non-Linear Schr\"odinger (NLS) equation. The thermodynamic Bethe ansatz structure of NLS has been considered in detail in Ref.~\cite{del_Vecchio_del_Vecchio_2020}. 

The limit of the zero temperature condensate in the NLS is particularly tricky. One of the best way to proceed is to take the $c\to 0^+$ limit of the Lieb-Liniger dressing and express the result in terms of principal value integral $\text{p.v.}\int\equiv\fint$, 
\begin{align}\label{eq:expansionlowc}
  &   \int_{-\theta_F}^{\theta_F}\frac{d\alpha}{2\pi}T(\theta-\alpha) f_{[n_{\rm LDA}]}^{\rm dr}(\alpha)  \nonumber \\& = f_{[n_{\rm LDA}]}^{\rm dr}(\theta) - 2c \fint_{-\theta_F}^{\theta_F} \frac{d \alpha}{2\pi}\frac{\partial_{\alpha}f_{[n_{\rm LDA}]}^{\rm dr}(\alpha)}{\theta - \alpha}, 
\end{align}
so that the NLS dressing operation in the quasi-condensate becomes
\begin{equation}\label{eq:dressing_NLS-0}
    f(\theta) = 2c\fint_{-\theta_F}^{\theta_F} \frac{d \alpha}{2\pi}\frac{\partial_{\alpha} f_{[n_{\rm LDA}]}^{\rm dr}(\alpha)}{\theta - \alpha} .
\end{equation}
From this expression (see Appendix \ref{app:semi-classical}),  $f^{\rm dr}(\theta)$ can be explicitly determined. In particular, one finds the well known semi-circle distribution for the $1^{\rm dr}(\theta)$,  
\begin{equation}\label{eq:semicircle}
   1_{[n_{\rm LDA}]}^{\rm dr} (\theta) =  \frac{1}{c}\sqrt{\theta_F^2 - \theta^2}  + {\cal O}(c^{-\frac{1}{4}}), 
\end{equation}
 with Fermi edges $\pm\theta_F$ given in terms of the density $\rho$ as 
\begin{equation}\label{eq:nls-fermi-edges}
 \rho(x) = \int \frac{d \theta}{2\pi} 1_{[n_{\rm LDA}]}^{\rm dr}(\theta)  =  \frac{(\theta_F(x))^2}{4c}      = \frac{\mu - V(x)}{c}.
\end{equation}
This expression agrees with the ground state density compute in LDA from the GP equation \eqref{eq:gross_pita}: in the bulk the gradient term can be put to zero giving \eqref{eq:nls-fermi-edges}.
As explicitly derived below, the zero coupling limit of $n_{\rm Fluct}$ yields the semi-classical ripples only. In this case, one can expand the phase in eq.~\eqref{eq:nFluct-LL} around the Fermi surface as done in Sec.~\ref{sec:hydro-evo}. Our starting point in assessing the zero coupling limit of $n_{\rm Fluct}$ is therefore the calculation of $\theta^{\rm dr}(\theta_F)$ for the quasicondensate. A simple calculation, based on Tricomi formula to invert the dressing operation \eqref{eq:dressing_NLS-0}, gives
\begin{equation}\label{eq:k-dress-NLS}
    \theta_{[n_{\rm LDA}]}^{\rm dr}(\theta)  = \frac{\theta}{2}   1_{[n_{\rm LDA}]}^{\rm dr}(\theta)
\end{equation}
implying that the zero-temperature effective velocity $v_{[n_{\rm LDA}]}^{\rm eff}(\theta)= \theta/2$, half of the bare one, for the $\theta$ lying inside the Fermi points, and $v_{[n_{\rm LDA}]}^{\rm eff}(\theta)= \theta$ for the $\theta$ outside. Now, the problem is that eq.~\eqref{eq:k-dress-NLS} together with eq.~\eqref{eq:semicircle} implies $\theta^{\rm dr}(\theta_F)=0$ in a non-analytic fashion. The behavior of dressed functions near the Fermi edges is hard to access in the low-coupling expansion as it requires to go beyond the leading terms in the expansion \eqref{eq:expansionlowc} at low $c$. We can however circumvent this problem by using the identity derived in Appendix \ref{app:great-identity}, giving this way the result
\begin{eqnarray}\label{eq:smallc1dr}
    1_{[n_{\rm LDA}]}^{\rm dr}(\theta_F) =  \sqrt{\pi} \left(\frac{\rho}{c}\right)^{1/4}  + \ldots
\end{eqnarray}
which, given that $ 1^{\rm dr}(\theta_F) = \sqrt{K}$, this implies ${K} =\pi/\gamma^{1/2} $ at low coupling as known from Bogoliubov theory~\cite{popov_theory_1977}. Eq. \eqref{eq:smallc1dr} implies, using eqs.~\eqref{eq:semicircle} and \eqref{eq:k-dress-NLS}, the following form for the dressed momentum at the Fermi point
\begin{equation}\label{eq:thetadrF}
  \frac{\pi \rho}{\sqrt{K}} =    \sqrt{\pi }
 \rho^{3/4} c^{1/4}  = \sqrt{\pi/c}    \  \rho_{\rm NLS}^{3/4}, 
\end{equation}
where, in order to take the limit to the classical NLS evolution, we have defined the rescaled NLS density in the limit $c \to 0^+$  
\be\label{eq:NLSScaling}
  \rho =  \rho_{\rm NLS}/c , 
\ee
such that the Fermi edges $\theta_F = 2 \sqrt{\rho c } = 2 \sqrt{\rho_{\rm NLS}}$ have a finite value in the limit $c \to 0$. Starting from eq.~\eqref{eq:nFluct-LL}, using $\theta^{\rm dr}_F(x)\equiv \theta^{\rm dr}(x,\theta_F(x))$ in \eqref{eq:thetadrF}, we obtain at weak interactions the fluctuating state
\begin{align}\label{eq:nfluct-NLS}
    n_{\rm Fluct} (x,k)  & =  \int dy\  {e^{\ii y  k } }  \nonumber \\& \times \frac{\sin\left( \sqrt{\frac{\pi}{c}}  \int_{x-y/2}^{x+y/2} ( \mu - V(x'))^{3/4}    dx' \right)}{\pi y}.
\end{align}
Within the approximation where $p^{\rm Dr}$ is computed using the LDA dressing (valid when $n_{\rm Fluct}$ is a small modification of the LDA filling), in the low coupling limit the mapping between $k$ and $\theta$ can be computed exactly using that $\phi(\theta)/(2\pi) \to   {\rm sgn}(\theta)/2$, giving this way 
\begin{eqnarray} 
    p_{[n]}^{\rm Dr}(\theta) \simeq \theta +   \frac{1}{2 c} \int_{-\theta_F}^{\theta_F}  d \alpha \ {\rm sgn}(\theta - \alpha)   \sqrt{\theta^2_F - \alpha^2} .
\end{eqnarray}
Performing the integral and using that $\sqrt{K} =  \rho_{\rm NLS}^{1/4}  \sqrt{\pi/c}$, the mapping reads 
\begin{align}\label{label:eqL}
   & \sqrt{c} \ k (\theta; x) =   \sqrt{\frac{1}{2\theta_F(x)}}    \Big( 2 c \ \theta + 
    \theta \sqrt{ (\theta_F(x))^2 - \theta^2}  \nonumber \\& + (\theta_F(x))^2 \arctan \left(\frac{\theta}{\sqrt{(\theta_F(x))^2 - \theta^2}} \right)    \Big),  
\end{align}
for $\theta \in [-\theta_F,\theta_F]$ and $ k (\theta; x) =   k (\pm\theta_F; x) $ for $\theta \gtrless \pm\theta_F$.  \\

First, we notice how in the homogeneous case $V(x)=0$, the integral over $y$ gives a unit step function for $ \theta^{\rm dr}(-\theta_F) \leq k\leq      \theta^{\rm dr}(\theta_F) $, hence reproducing the LDA filling. Moreover, by scaling $ y \to \sqrt{c} y$, we notice that the integral over the phase $y$ in eq. \eqref{eq:nfluct-NLS} acts only on a region $\sim {\cal O}(\sqrt{c})$ and therefore for small coupling we can write 
\begin{align} \label{eq:finalNLSS}
  &   n_{\rm Fluct} (x,\theta)   =  \int dy\  {e^{\ii y  \sqrt{c} k(\theta; x) } }  \nonumber \\& \times \frac{\sin\left(   ( \mu - V(x))^{3/4}  y  + \frac{y^3 c }{24} \partial_x^2  ( \mu - V(x))^{3/4} + O(c^2)  \right)   }{\pi y}.
\end{align}
 We obtain again the result of eq.~\eqref{eq:nfluctAiry}, now with  $\alpha \sim c$, giving therefore a smearing of the $\delta$ functions in the zero-entropy GHD proportional to $c^{1/3}$. Notice that the $c$ correction to the phase is the same order as the $c$ correction to $k$ in eq. \eqref{label:eqL}. It is therefore evident that the density ripples at small coupling have a semi-classical nature. Taking the limit $c\to 0 $ in eq. \eqref{eq:finalNLSS}, we finally find that in the strict limit of the classical NLS, the effects of the initial quantum fluctuations vanishes, as expected, and quantum ripples, {as the one carried by $n_{\rm Fluct}$}, amount to subleading corrections, see Appendix~\ref{app:qghd-nls}. On the other hand we remark that  NLS equation is known to have density ripples caused by the dispersive wave breaking, as  extensively discussed in the literature using Whitham approach, see e.g.~\cite{whitham_1974,forest_1986,pavlov1987nonlinear,el_dispersive_2016}. It is clear that such ripples therefore have a different origin, they must come from dispersive terms in the GHD evolution, see eq. \eqref{eq:EulerGHD2}. Their study and a more direct comparison of our approach with Whitham theory is postponed to subsequent studies. 

\section{Numerical results}\label{sec:discuss}
We now move to the numerical simulations of the evolution of $n_{\rm Fluct}$ via Euler GHD equation eq. \eqref{eq:EulerGHD}.\\
Prepared in an initial density hump, Fig.~\ref{fig:isometric_density} compares the evolution of the density profiles given by $n_{\rm fluct}$ for two cases: for $\gamma = 1$ (panel~(a)) and for $\gamma = 100$ (panel~(b)), noting that the latter is far into the Tonks-Girardeau regime. In Fig.~\ref{fig:isometric_density}(b), the analytical calculations for the Tonks-Girardeau gas of Sec.~\ref{sec:tonks} allow us to distinguish between the small wavelength oscillations, or quantum fluctuations, and longer wavelength semi-classical corrections. Away from this limit any clear separation of these corrections is substantially more challenging. However, by observation, the leading corrections to the LDA density in Fig.~\ref{fig:isometric_density}(a) appear to be mostly those corresponding to long wavelength corrections as $\gamma$ is decreased. These long wavelength corrections are found most visibly at the edges of the shock region, similarly to the long wavelength semi-classical corrections of Fig.~\ref{fig:isometric_density}(b). Further corroboration of the semi-classical origin for these corrections is the small $c$ expansion of the $n_{\rm fluct}$ in the GPE limit, (see Sec.~\ref{section:semi-classical_limit}), whose leading order correction is found to exhibit an analogous form as those semi-classical ripples in the Tonks-Girardeau limit. Altogether, these observations suggest that the small and long wavelength corrections to LDA originate from distinct mechanisms, and that the semi-classical corrections persist into the GPE limit.

In Fig.~\ref{fig:nfluct-gammas}(a), we plot a phase-space color map of $n_{\rm Fluct}$ for different values of $\gamma$ (increasing from top to bottom) and times (increasing from left to right). Fig.~\ref{fig:nfluct-gammas}(b) displays a cross-section of the initial fluctuating state at $x=0$ as function of $\theta$, emphasizing its deviations from a perfect Fermi sea. Starting from the bottom row (corresponding to the Tonks regime $\gamma\gg 1$), we see that $n_{\rm Fluct}(t=0; x,\theta)$ has several stripes near the Fermi edges (red curves), indicating the initial correlations present in this fluctuating state. Similar patterns have been previously observed and can be understood as contours of constant semi-classical action \eqref{eq:semiclass-action}, see Ref.~\cite{Protopopov2013,protopopov2014}.\\
 On decreasing the interaction to finite values, these patterns undergo substantial changes, qualitatively differing from the LDA envelope (red curves) calculated using zero-entropy GHD \eqref{eq:zeroEGHD}. In particular, we observe that the finer structures at $\gamma\gg 1$ group together to form large-scale oscillations (see also the cross-sections in Fig.~\ref{fig:nfluct-gammas}(b)), and spread much further away from the Fermi edges. Explaining these oscillations in terms of saddle points of the semi-classical action~--~as in the Tonks-Girardeau regime~--~ is challenging at finite interactions. Nevertheless, a qualitative explanation of these fringes can still be found by looking at the phase in eq.~\eqref{eq:nFluct-LL}, $d\Phi_\text{int}/dx=\pi\rho_\text{LDA}/\sqrt{K}$ for a fixed density profile $\rho_\text{LDA}$: while fine structures at $\gamma\gg 1$ correspond to high-frequency quantum ripples similarly to Friedel oscillations in boundary systems, when $\gamma$ is decreased $K$ monotonically increases \cite{Cazalilla2004,Cazalilla2011} and thus the scaling factor $\sqrt{K}$ enhances long wavelength modulations. This is consistent with standard bosonization arguments, for which Friedel-like oscillations in the shock region are expected to decay as $\sim 1/|x|^K$, thus vanishing when $K$ diverges~\footnote{The non-universal amplitude $A(\gamma)$ of this process drops quicky to zero when $\gamma$ is decreased from the Tonks-Girardeau limit, enforcing the disappearance of Friedel-like oscillations.}.
 \\
 
 Fig.~\ref{fig:density-gammas-DMRG} shows the evolution of particle density for the setup of Figure~\ref{fig:shock}. Specifically, the top row displays the dynamics at different $\gamma$ as predicted by $n_{\rm Fluct}$, while the bottom row shows the corresponding evolution obtained numerically. Simulating the Lieb-Liniger gas \eqref{eq:lieb-liniger-H} is notoriously challenging, particularly for out-of-equilibrium analysis where MonteCarlo methods are usually inapplicable. In Fig.~\ref{fig:density-gammas-DMRG}, we simulate the Lieb-Liniger gas exploiting its mapping to the XXZ spin chain at small densities (as done e.g. in Refs.~\cite{Brun2018,Peotta2014,Riggio2022,Schmidt2007} and discussed in Refs.~\cite{Golzer1987,Pozsgay2011}), employing time-dependent Density-Matrix Renormalization Group (tDMRG) on this lattice model. 
To access large times $t\gtrsim t_s$, we consider open boundary conditions on the spin chain, which induce visible Friedel oscillations in the background, especially in the limit of large $\gamma$. Despite these limitations, we have identified a parameter set enabling the simulation of the post-shock dynamics of the Lieb-Liniger gas. We are aware of other methods, e.g. infinite Matrix Product States algorithms, which may offer more accurate numerical analysis; however, these are deferred to subsequent studies, possibly involving experts in such numerical techniques.\\
 
 In our opinion, Fig.~\ref{fig:density-gammas-DMRG} unequivocally validates our approach based on the fluctuating initial state for finite interactions. Indeed, for each value of $\gamma$ and time, our method accurately captures the shape, the number, and the location atop the LDA solution of the density ripples. We observe some quantitative deviations from the tDMRG data at small values of interaction, for instance, at $\gamma=2$ shown in the plot. Although we conducted our numerics for density humps with larger variance (within our computational capabilities), we still note slight differences between the initial state in the DMRG and the LDA result, thereby introducing errors in the observed quantum fluctuations after the shock. Corrections to LDA, as discussed for example in Ref.~\cite{Riggio2022}, may be considered for future improvements. These corrections are intimately connected to higher-order derivative terms in the hydrodynamic evolution. Indeed, in general, one has dispersive corrections also in the LDA evolution,  namely the Euler-scale hydrodynamics \eqref{eq:EulerGHD} is extended by 
\begin{align}\label{eq:EulerGHD2}
\de_t &  n_{\rm Fluct}  + v^{\rm eff}_{[n_{\rm Fluct}]}  \de_x n_{\rm Fluct} 
= \partial_x (\mathcal{W}_{[n_{\rm Fluct}]} \cdot \partial^2_x n_{\rm Fluct}) ,
\end{align}
where the dispersive kernel for the Lieb-Liniger gas has been introduced in \cite{denardis_doyon_dispersive}. Such classical dispersive terms vanish in the strong coupling limit as $1/c^3$ but they become more important in the limit $c \to 0^+$, giving this way additional oscillations, of different natures compared to the quantum fluctuations introduced by $n_{\rm Fluct}$. We leave the question of including these terms to future work.

\section{Conclusions and outlooks}\label{sec:conclusion}

In this paper we have provided a unified theory for density ripples and wave breaking effects in the one-dimensional Bose gas, by combining the hydrodynamic evolution and initial quantum fluctuations. We have shown that typical density ripples observed in density waves propagation emerges from quantum fluctuations of the initial state, which can be characterized in terms of effective fermionic excitations around the Fermi seas. By evolving the initial Wigner function for such fermionic degrees of freedom with Euler GHD evolution, we are able to compute the quantum correction to the hydrodynamics, or LDA, mean density.

Our work open numerous different directions. First,  our approach based on $n_{\rm Fluct}$ can be directly applied and extended to different experimentally relevant settings, such harmonically trapped bosons or domain walls in spin chains.  Moreover, it would be desirable to benchmark it with the more standards approaches of quantum GHD and Luttinger liquid, based on bosonized excitations \cite{Ruggiero2020}. Given the new approach to Luttinger liquid correlations that our work employs, it is reasonable to expect that such comparison will shine some lights in different aspects of the connection between Luttinger liquid and Bose gas, as for example the Luttinger liquids' prefactors for the static and dynamic correlation functions~\cite{PhysRevB.85.155136,https://doi.org/10.48550/arxiv.1110.0803}, for which a thermodynamic closed-form expression is still out of reach. 
Moreover, a more precise connection with bosonization will be important to understand how to extend our formalism to compute observables beyond one-point functions of local densities, as two-point functions~\cite{Ruggiero2019,Scopa2023,takacs2024quasicondensation, PhysRevLett.131.263401, Del_Vecchio_Del_Vecchio_2022} and entaglement entropy~\cite{Scopa2021,Collura2020} .

The inclusion of higher-order derivatives to the Euler GHD evolution of $n_{\rm Fluct}$ also represent a challenging topic for the near future, and in particular the strictly classical limit to the NLS equation.  Moreover it would be tempting to extend known concept of classical diffusion, to the quantum case.  In integrable models indeed diffusion is manifested by classical, finite temperature, fluctuations due to random density waves \cite{10.21468/SciPostPhys.6.4.049,gopalakrishnan2020hydrodynamics,10.21468/SciPostPhys.9.5.075,2401.05494}. At zero temperature such thermal fluctuations clearly vanish but quantum ones are instead present here inhomogenous cases: while $n_{\rm LDA}$ is a zero-entropy distribution (namely taking only $0$ and $1$ values), this is not the case for $n_{\rm Fluct}$, allowing therefore for diffusive-like fluctuations. How to extend the formalism of GHD to include such quantum diffusive terms and quantum fluctuating terms to access large deviation functions \cite{PhysRevB.109.024417,PhysRevLett.128.090604,2401.05494},  is an exciting quest for the near future. 
\\

\prlsection{Acknowledgements} We acknowledge M. Fagotti, I. Bouchoule, F. Essler, K. Kheruntsyan and E. Bettelheim for useful discussions. We in particular acknowledge J. Dubail, P. Ruggiero, B. Doyon and T. Bonnemain for discussions and early collaborations on similar subjects. The authors are grateful to J. Dubail for sharing unpublished results  (Ref.~\cite{Scopa-Dubail-tonksripples}) and for critical remarks on this  project. This work has been partially funded by the ERC Starting Grant 101042293 (HEPIQ) (J.D.N., S.S and A.U.) and ANR-22-CPJ1-0021-01 (J.D.N). Tensor network simulations have been performed using iTensor library~\cite{itensor}. This work was granted access
to the HPC resources of IDRIS under the allocation
AD010513967R1 and A0140914149.

\section*{Appendixes}
\appendix
\section{Low-energy description of the excitations}\label{app:bosonization}
This appendix serves as a dictionary of known results from bosonization theory. For a detailed discussion on the following equations, see e.g. Refs.~\cite{Giamarchi2003,Cazalilla2004,Gogolin1998,Tsvelik2007,Senechal1999}.\\

To begin with, we consider the Haldane's harmonic fluid expansion \cite{Haldane1981,Haldane1981a} of the (bosonic) density operator
\begin{align}\label{eq:harmonic-exp}
\hat\rho(x)=&\left(\rho_0+\frac{\de\hat{\bm{\varphi}}(x)}{\pi}\right) \sum_{m=-\infty}^{\infty} A_m :e^{2\ii m(k_F x + \hat{\bm\varphi}(x))}:\nn \\
&\simeq  \rho_0 +\frac{\de\hat{\bm{\varphi}}(x)}{\pi} +\rho_0(A_1 :e^{2\ii(k_F x + \hat{\bm{\varphi}}(x))}: \nn \\
&\qquad +A_{-1} :e^{-2\ii (k_F x + {\bm{\varphi}}(x))}:).
\end{align}
 $k_F\equiv p^\text{Dr}(\theta_F)=\pi\rho_0$ here denotes the Fermi momentum of the interacting gas; $\hat{\bm\varphi}(x)$ is the bosonic fluctuating field (also called height field), which can be decomposed into its chiral components as
\be
\hat{\bm\varphi}(x)=\frac{\sqrt{K}}{2}\left(\hat\varphi_+(x)+\hat\varphi_-(x)\right),
\ee
and the latter have two-point correlations
\be\label{eq:chiral-2point}
\langle \hat\varphi_\sigma(x) \hat\varphi_{\sigma'}(x')\rangle=\delta_{\sigma,\sigma'}\log(\sigma \ii (x-x').
\ee
Next, we write the density operator in \eqref{eq:harmonic-exp} as 
\be
\hat\rho(x)=\rho_0 + \delta\hat\rho^\text{long}(x)+\delta\hat\rho^\text{short}(x).
\ee
The first term, $\delta\hat\rho^\text{long}=\de\hat{\bm\varphi}/\pi$, contains the long wavelength modulation of the density, occurring for those low-energy processes that take place around the two Fermi points (particle-hole pairs). Pictorially,
\be
\delta\hat\rho^\text{long}:\quad \includegraphics[scale=0.4]{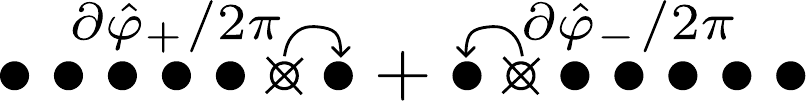}
\ee
with amplitude $\langle\text{part.-hole}|\delta\hat\rho(0)|\text{g.s.}\rangle/L=\sqrt{K}$ in the thermodynamic limit \cite{DeNardis2015,Bouchoule2023}. The second term, 
\be\label{eq:Umklapp}
\delta\hat\rho^\text{short}= \rho_0 A_{\pm1} (:e^{2\ii (k_Fx+\hat{\bm\varphi}(x))}: + :e^{-2\ii (k_Fx+\hat{\bm\varphi}(x))}:)
\ee
contains the Umklapp processes at lowest order
\be
\delta\hat\rho^\text{short}: \quad \includegraphics[scale=0.4]{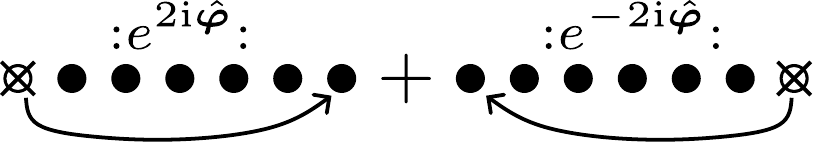}\; .
\ee
The non-universal amplitudes $A_m(\gamma)$ are extracted from the form factor of the density operator, $\langle \text{Umklapp}(m)|\hat\rho(0)|\text{GS}\rangle \equiv A_m$. At lowest order, $A_{1}=A_{-1}$ see e.g. Ref.~\cite{Brun2018} for its plot as function of $\gamma$.

At this point, we consider an alternative description of these excitations in terms of fermionic field, similarly to what has been done in the main text. We introduce the following fermionic operators
\begin{align}\label{eq:refermionized}
&\hat{c}^\dagger(x)= \hat\psi_+^\dagger(x) + \hat\psi_-^\dagger(x)\\
&\hat{c}(x)=\hat\psi_+(x)+ \hat\psi_-(x)
\end{align}
with chiral components
\begin{align}
&\hat\psi_+(x)= \frac{\eta  e^{-\ii k_F x}}{\sqrt{2\pi}}  :e^{-\ii \hat\varphi_+(x)}:,  \quad \hat\psi_-(x)= \frac{\bar\eta e^{\ii k_F x}}{\sqrt{2\pi}}  :e^{\ii \hat\varphi_-(x)}:
\end{align}
where $\{\eta,\bar{\eta}\}=0$, $\eta^2=\bar{\eta}^2=1$ are Klein factors ensuring anticommutations of the chiral fermions in their bosonic representation, $\{\hat\psi_+(x),\hat\psi_-(y)\}=0$, $\{\hat\psi_+^\dagger(x),\hat\psi_+(y)\}=\delta(x-y)$ and so on. Usually, these are represented with Pauli matrices $\eta=\sigma^x$, $\bar\eta=\sigma^y$. Since they are unimportant in what follows, we shall omit these factors in the discussion below. The factor $1/\sqrt{2\pi}$ is here inserted for convenience. \\

 The fermionic propagator is easily calculated as
\begin{align}
&{\cal K}(x,y)=\langle \hat{c}^\dagger(x)\hat{c}(y)\rangle\\
\nonumber
&=\langle \hat\psi_+^\dagger(x)\hat\psi_+(y)\rangle + \langle \hat\psi_-^\dagger(x)\hat\psi_-(y)\rangle=\frac{\sin(k_F(x-y))}{\pi(x-y)},
\end{align}
and for $x\to y$ we recover the value of density $\rho_0=k_F/\pi$.\\

We remark that this description does not account for the Umklapp processes that mix the two chiralities. In fact, defining the fermionic density operator as
\be
\hat\rho_F- \rho_0=\hat{c}^\dagger(x)\hat{c}(x),
\ee
the shortwavelenght contribution in eq.~\eqref{eq:Umklapp} reads as
\be
 \delta\hat\rho_F^\text{short}(x)=\hat\psi_+^\dagger(x)\hat\psi_-(x)+\hat\psi_-^\dagger(x)\hat\psi_+(x),
\ee
and it cannot be simply expressed in terms of its bosonic counterpart $\delta\rho^\text{short}(x)$. The reason behind this limitation is the non-universality of the process described by \eqref{eq:Umklapp}. On the other hand, the fermionic long wavelength contribution is 
\be
\delta\hat\rho_F^\text{long}(x)=\hat\psi_+^\dagger(x)\hat\psi_+(x) + \hat\psi_-^\dagger(x)\hat\psi_-(x) 
\ee
Such terms are ill-defined and must be computed using point splitting
\begin{align}
&\hat\psi^\dagger_+(x)\hat\psi_+(x)=\lim_{\epsilon\to0}\bigg(\hat\psi^\dagger_+(x+\epsilon)\hat\psi_+(x-\epsilon)\nn \\
&-\langle\hat\psi_+^\dagger(x+\epsilon)\hat\psi_+(x-\epsilon)\rangle \bigg) = \frac{\de\hat\varphi_+(x)}{2\pi},
\end{align}
and the same for the left part. Therefore,
\be
\delta\hat\rho_F^\text{long}(x)=\frac{\de\hat{\bm\varphi}(x)}{\pi\sqrt{K}} =\frac{\delta\hat\rho^\text{long}(x)}{\sqrt{K}},
\ee
which is indeed eq.~\eqref{eq:density-exp} of the main text.
\section{Calculation of post-shock density in the Tonks-Girardeau regime}\label{app:correction-lda-tg}
In this appendix, we detail the calculation of density ripples in the Tonks-Girardeau regime.\\

We start by writing $s(\vec{x})$ in eq.~\eqref{eq:s-coo} as
\be
s(\vec{x})=\begin{dcases}
    \int_{-r}^x \frac{dx'}{p_F(x')}, \quad\text{if $\theta>0$}; \\
    2{\cal L}-\int_{-r}^x \frac{dx'}{p_F(x')}, \quad\text{if $\theta<0$},
\end{dcases}
\ee
${\cal L}=\int_{-r}^r dx'/p_F(x')$, where $r$ is a reference point located away from the external perturbation, $V(\pm r)=0$. Its specific value is unimportant since it drops out from the calculation of physical quantities. We mention that this parametrization along the contour can be related to a WKB quantization condition on the phase space,
\begin{eqnarray}
    2 \pi N_r= \oint ds \ \theta_0(s),
\end{eqnarray}
with $N_r$ being the number of particles in the system in the region $[-r,r]$. In addition, the upper- ($\theta>0$) and lower- ($\theta <0$) half of the $x$-$\theta$ plane (corresponding to the two chiralities) decouple and equally contribute to the phase-space quantization condition above. Hence, without loss of generality, one can focus on the upper contour and use \eqref{eq:s-coo} as definition of the stretched coordinate $s(\vec{x})\equiv s(x)$.\\

Next, we notice that the time-evolved two-point function
\be
{\cal K}_\pm(t; x,x')=\langle V| e^{\ii t \hat{H}}\, \hat\psi^\dagger_\pm(x)\hat\psi_\pm(x') \, e^{-\ii t \hat{H}} |V\rangle
\ee
 is unambiguously defined in terms of $s(x)$ and reads as
\be
{\cal K}_\pm(x,x')=\mp \ii\sum_{\{x_0,x_0'\}}\sqrt{\left|\frac{d s({x_0})}{d x} \right|\left|\frac{d s({x_0'})}{d x}\right|} \frac{e^{\pm\ii \Delta\Phi_t(x_0,x_0')}}{s(x_0)-s(x_0')}.
\ee
Indeed, given the values of $x$ and $t$, there might be multiple points $s_a \in \Gamma_0(s)$ such that $x_{0}(s_a)=x-\theta(s_a) t$.\\

At this point, we are ready to focus on the calculation of particle density. Specifying to the points $s_{1,2}\equiv s_{1,2}(\theta, y)$ such that $x_0(s_{1,2})=x-t\theta \mp y/2$, it can be defined as
\begin{align}\label{eq:dens-start}
    \langle \hat\rho(t,x) \rangle =\iint &\frac{dy d\theta \, e^{\ii \theta y} }{2\pi^2} \sqrt{\left|\frac{d s_1}{d x} \right|\left|\frac{d s_2}{d x}\right|} \nn \\
    &\times\, \frac{\sin (\Delta\Phi_t[x_0(s_1),x_0(s_2)])}{s_2 - s_1}.
    \end{align}
Introducing the semi-classical action \eqref{eq:semiclass-action} $S(\theta,y)=-\theta y  +\Delta\Phi_t$, we look for the saddle point $(\theta,y)$
\be\begin{dcases}
&\frac{\de S(\theta,y)}{\de \theta}\bigg\vert_{y\to 0}=0;\\[3pt]
&\frac{\de S(\theta,y)}{\de y}\bigg\vert_{y\to 0; \theta\to \frac{dx_t}{dx_0} \times p_F(x_0)}=0,
\end{dcases}
\ee
implying that the main contribution to eq.~\eqref{eq:dens-start} is coming from $y\to 0$ 
 \begin{align}\label{eq:starting-point}
    \langle\hat\rho(t,x) \rangle =\sum_{\{s_{1,2}\}} \int& dy  \frac{\delta (y)}{2\pi}  \sqrt{\left|\frac{d s_1}{d x} \right|\left|\frac{d s_2}{d x}\right|} \nn\\
    &\frac{\sin (\Delta\Phi_t[x_0(s_1), x_0(s_2)])}{s_2 - s_1}.
\end{align}   
 The set $\{s_{1,2}\}$ in the sum contains the coordinates satisfying 
 \be
 \{s_{1,2}\} =\left\{ s \in \Gamma_0(s) \ : \ x_0(s)=x\mp y/2-\theta_0(s) t\right\},
 \ee
eventually becoming two indenpendent solutions of $x_0(s)=x-\theta_0(s) t$ when $y$ is set to $0$. The jacobians are easily evaluated starting from eqs.~\eqref{eq:eq-of-motion} and \eqref{eq:s-coo}, reading (cf. eq.~\eqref{eq:jacobians} of the main text)
 \be
 \frac{d x_t(s)}{ds}=\theta_0(s)\Big(1 + t \ p'_F[x_0(s)]\Big).
 \ee
  Three different regions need to be distinguished: \emph{i) LDA region}, $x < x_-(t)$ and $x > x_+(t)$; \emph{ii) shock region}, $x_-(t) < x < x_+(t)$; and \emph{iii) caustic points}, $x \approx x_\pm(t)$. 
\\

 Here,  $x_\pm(t)$ are determined by eq.~\eqref{eq:shock-points}, which we rewrite as
\be
\frac{d x_t(s)}{d s}\bigg\vert_{s={s}_\pm}=0, \quad x_\pm(t)\equiv x_t(s_\pm)
\ee
and we shall refer to ${s}_\pm$ as turning points on the contour.\\

\begin{figure}
\centering
\includegraphics[width=.7\columnwidth]{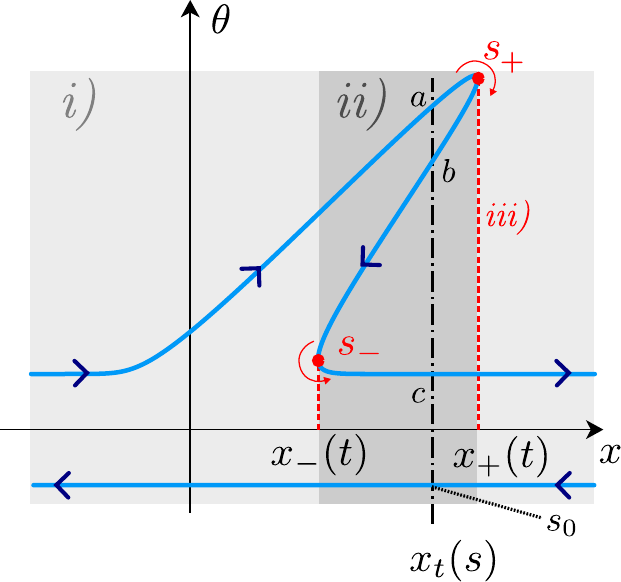}
\caption{Illustration of the Fermi contour for $t>t_s$. }\label{fig:ps-contour}
\end{figure}

 We illustrate the phase-space configuration during the post-shock dynamics in Figure~\ref{fig:ps-contour}. Notice that regardless of the region, one always finds the solution $s_0\in \{s\}$ which amounts for half of the background contribution in the initial state, $\theta_0(s_0)/(2\pi)$, to the particle density. Therefore, below we consider $\langle\delta\hat\rho\rangle\equiv \langle\hat\rho\rangle-\theta_0(s_0)/(2\pi)$, i.e. we exclude $s_0$ from the set of coordinates in the sum of  \eqref{eq:starting-point}: $\{s\}_\star\equiv \{s\}\setminus s_0$.\\

\subsection*{\emph{i)}~LDA region, $x < x_-$ and $x > x_+$}
In this region, there is only one element in $\{s\}_\star$ and it is located away from the turning points. The limit $y\to 0$ yields $s_2-s_1\simeq y  \de_x s$ and $\Delta\Phi_t[x_0(s_1), x_0(s_2)] \simeq p_F(x_0(s)) y$. Substituting this in \eqref{eq:starting-point}, we obtain
\begin{eqnarray}
    \langle \delta\hat\rho \rangle = \frac{ \theta_0(s)}{2\pi} .
\end{eqnarray}
Note that this is exactly the expected LDA result.\\

\subsection*{\emph{ii)}~Shock region, $x_- < x < x_+$}
In this case, there are three distinct roots $s_a<s_b<s_c$ in $\{s\}_\star$ away from the turning points, see Fig.~\ref{fig:ps-contour}. For each of these, selecting $s_1=s_2$ in \eqref{eq:starting-point} one obtains the LDA contribution in the shock region (albeit with two Fermi seas)
\begin{eqnarray}
    \langle \delta\hat\rho \rangle_{\rm LDA} =\frac{\theta_0(s_a)-\theta_0(s_b)+\theta_0(s_c)}{2\pi}.
\end{eqnarray}
The minus sign in the second term comes from the fact that $\de_x s<0$ on the middle brach of the contour.\\

 In addition to such LDA terms, multiple solutions in $\{s\}_\star$ allow for ``non-diagonal'' contributions $s_1\neq s_2$. These terms physically correspond to the quantum interference of two Fermi points, and generate the density ripples over the LDA result. Notably, this mechanism involves only the Fermi points sitting along the upper contour (resp. lower contour for the left-moving part). Indeed, the interference with $s_0$ would require traveling for an infinite distance along the contour. Specifically there are 3 contributions, $(i,j)=(a,b)$, $(a,c)$ and $(b,c)$. Each of these gives,
\begin{align}
    \langle \delta\hat\rho\rangle\bigg|_{i,j} &\approx  \sqrt{\left|\frac{d s_i}{dx}\right|\left|\frac{d s_j}{dx}\right|} \frac{ \sin\left(\lim_{y\to0} \Delta\Phi_t[x_0(s_i),x_0(s_j)] \right)}{2\pi(s_i-s_j)} \, . \nn\\
\end{align}
The additional complication here is in determining of the phase $\lim_{y\to 0}\Delta\Phi_t[x_0(s_i),x_0(s_j)]$. Indeed, it should be noted that the integral in eq.~\eqref{eq:time-dep-WKB}, which can be also written as
\be
\Delta\Phi_t(s_1,s_2)=\int_{s_1}^{s_2} dx_t(s) \;\theta_0(s),
\ee
now involves passing over turning points $s_\pm$ along the path from $s_i\to s_j$, see also Fig.~\ref{fig:ps-contour}. This feature must be accounted for when mapping the system to its initial state. A possibility is to introduce Maslov indices~\cite{Maslov_1981}, defined as \emph{${\cal I}_{ij}\equiv$ number of counterclockwise turning points $-$ number of clockwise turning points along the path $s_i\to s_j$}. Using this, we obtain
\begin{align}\label{eq-maslov}
    \langle \delta\hat\rho\rangle\bigg|_{i,j} &\approx  \sqrt{\left|\frac{d s_i}{dx}\right|\left|\frac{d s_j}{dx}\right|} \frac{ \sin\left( \Delta\Phi_t[x_0(s_i),x_0(s_j)]  - {\cal I}_{i j} \frac{\pi}{2}\right)}{2\pi(s_i-s_j)} \, . \nn\\
\end{align}

The phase shift involving Maslov indices in \eqref{eq-maslov} is physically interpreted as a boundary point reflecting the chiral modes along the contour. At this point, the final step is to sum over all the $(i,j)$ pairs and to determine the corresponding Maslov indices. The last task is easily done:
\be
{\cal I}_{ij}=\begin{cases}
-1,\quad \text{if $(i,j)=(a,b)$}\\
+1,\quad\text{if $(i,j)=(b,c)$}\\
0,\quad\text{if $(i,j)=(a,c)$},
\end{cases}
\ee
then, defing  $J_{ij}= \sqrt{\left| (d s_i/d x)(d s_j/d x )\right|}$, the sum of all terms gives eq.~\eqref{eq:qghd_ripples} of the main text.

\subsection*{\emph{iii)} Caustic points, $ x \approx x_\pm$}
Finally, we discuss the corrections to LDA density in the region $x\approx x_\pm(t)$. Close to these points, the Fermi contour develops what are known as  \emph{caustics}, i.e. points where $|d\theta_0(s)/dx_t(s)|_{s=s_\pm}=\infty$. Despite this complication, many methods for studying these regions have been developed~\cite{berry_semi-classical_1997,arnold1978}. For simplicity, below we focus on point $x_+$, although the analysis of $x_-$ proceeds analogously. \\

Close to the caustic $x_+$, the top and the middle branches of the Fermi contour connect, and two roots $s_b\to s_a$ coincide, see Fig.~\ref{fig:ps-contour}. At this point, starting from eq.~\eqref{eq:starting-point} one finds as above the LDA contribution
\be
\langle\delta\hat\rho(x\approx x_+)\rangle_\text{LDA} =\frac{\theta_0(s_c)}{2\pi},
\ee
and the two regular interference terms $(a,c)$ and $(b,c)$ given in eq.~\eqref{eq-maslov}. This leaves the contribution of $(a,b)$ as $s_b\to s_a$, where it becomes singular, that requires a separate treatment.\\

 By further expanding the phase in \eqref{eq:dens-start} around $y=0$ for $s\approx s_\pm$, one finds 
\begin{align}
\langle \hat\rho \rangle_{\text{UV}} &\approx \iint \frac{d\theta  dy}{(2\pi)^2 \ii y} \exp\bigg[\ii (\theta_0(s) - \theta ) y  + \ii\frac{p_F''(x_0(s))}{24}y^3\bigg]  \nn\\
 &= \int \frac{d\theta}{2\pi } \text{Ai}_1\left( \left[\frac{-8 \ii }{ p''_F(x_0(s))}\right]^{1/3} (\theta - \theta_0(s)) \right) ,
\end{align}
with the special function $\text{Ai}_1 (x) = \int dq/(2\pi \ii) \exp[\ii( q x +  q^3/3)]/q$.
 Alternatively, this expression may be written instead in terms of the Fermi position and by so doing find an expression which remains valid near the caustic point caustic point at time $t$:
\begin{align}\label{eq:bette-position}
\langle \hat\rho \rangle_{\text{UV}}  &\approx \int \frac{d\theta}{2\pi} \text{Ai}_1\left( \left[  \frac{\ii8}{x''_F(\theta_0(s'))} \right]^{1/3} (x_t(s') - x)  \right).
\end{align}
Here it is important to note that the free variable $x$ and the Fermi position $x_t(s')$ are distinct from one another. To obtain this result, consider the case where the free variable $\theta$ in the initial state lies near the Fermi contour such that there exists an $s'$ satisfying $\theta_0(s') \simeq \theta$. In this case the expansion
\begin{eqnarray}
  \theta_0(s) - \theta &\approx&  p_F'(x_0(s')) (x- x_t(s')) \nn\\
    &=& \frac{x - x_0(s')}{x_F(\theta_0(s'))},
\end{eqnarray}
with 
$ p_F'(x_0(s)) = 1/(x_F'(\theta_0(s))$. Differention of the latter identity with respect to the curve $s$ yields after some algebra
\begin{eqnarray}
    p_F''(x_0(s)) = - \frac{x_F''(\theta_0(s))}{(x'_F(\theta_0(s)))^3}.
\end{eqnarray} 
This identity is particularly useful when combined with eq.~\eqref{eq:eq-of-motion}, since it implies that $x_F''(\theta_0(s)) = x_F''(\theta_t(s))$, i.e. that the second derivative of the Fermi position is time independent!
Taken together, the last two permit the identification
\begin{align}
    \left[\frac{-\ii 8}{ p''_F(x_0(s))}\right]^{\frac 1 3} (\theta- \theta_0(s)) \approx (x_0(s') - x) \left[  \frac{\ii8}{ x_F''(\theta_0(s'))} \right]^{\frac 1 3},\nn \\
\end{align}
it might appear that nothing has been accomplished, but in this form the expansion around the caustic is now valid for any time. More concretely $x_t'(s^\pm) = 0$, so the small $y$ approximation is remains valid in the neighborhood of the caustic point, and $x''_F(\theta_0(s'))$ is time independent. So the initial Fermi position can be mapped onto the Fermi position at time $t$ by the identification $x_0(s') \to x_t(s')$ thus obtaining eq.~\eqref{eq:bette-position}.\\

All that remains now is to approximate the integral of eq.~\eqref{eq:bette-position}, which is most easily done by first taking the derivative with respect to the free variable $x$. From the definition of $\text{Ai}_1$ it follows that
\begin{align}\label{eq:deriv}
\partial_x \langle \hat\rho \rangle_\text{UV} &=  \iint \frac{d\theta dq }{2\pi}
 \exp\bigg[\ii (x- x_t(s')) q + \frac{\ii|x_F''(\theta_0(s'))|q^3}{24}\bigg],
\end{align}
with the $\theta$ dependence hidden in $x_t(s')$. For points close to the shock points at $s_\pm$ it is possible to expand $x_t(s')$ making the $\theta$ dependence explicit, then find (recall $x_F'=0$ at the caustic point) 
\begin{align}\label{eq:grad-exp}
     x_t(s')- x_\pm(t) \approx x''_F(\theta_0(s_\pm)) \frac{ p^2}{2},
\end{align}
where $p = \theta -\theta_0(s_\pm)$. Using \eqref{eq:grad-exp} in \eqref{eq:deriv} and changing coordinates to $P = p + \frac{q}{2}$ and $Q = p - \frac{q}{2}$, one finds upon substitution of these new variables that mixed terms cancel and the integrand becomes separable
\begin{align}
&\partial_x \langle \hat\rho \rangle_\text{UV} \approx  \iint \frac{dP dQ}{(2\pi)^2} e^{\ii (x- x_\pm(t)) (P-Q) + \ii\frac{|x_F''(\theta_0(s_\pm))|}{6}(P^3 - Q^3)} \nn\\
&=  \left\{\text{Ai}\left( \left[\frac{2 \ii }{ x_F''(\theta_0(s_\pm))} \right]^{1/3} (x- x_\pm(t))  \right) \right\}^2.
\end{align}
For brevity introducing $\kappa^\pm = \left[ 2 \ii / x_F''(\theta_0(s_+))\right]^{1/3}$, $\delta x_\pm= \mp (x-x_\pm)$, and so conclude that
\be
    \partial_x \langle\delta \hat\rho\rangle_{\text UV}  =  (\kappa^\pm)^2 \left[ \text{Ai}(\kappa^\pm \delta x_\pm) \right]^2,
    \ee
    which can be integrated to give eq.~\eqref{eq:caustic} of the main text. \\
    
With the technical steps concluded, we remark that the expectation value of density for $x\approx x_+(t)$ is thus given by the sum of the LDA terms, of the regular terms in \eqref{eq-maslov} and of \eqref{eq:caustic}, that is
\begin{align}
   & \langle \hat\rho(x\approx x_+(t))\rangle= \frac{\theta_0(s_c)}{2\pi} +J_{bc} \frac{\cos\left(\Delta\Phi_t[x_0(s(x)),x_0(s_c)]\right)}{\pi|s(x)-s_c|} \nn \\
   &-J_{ac}\frac{\sin\left(\Delta\Phi_t[x_0(s(x),x_0(s_c)]\right)}{\pi|s(x)-s_c|}\nn \\
    & + (\kappa^+)^2 \bigg[  \delta x_+ [\text{Ai}(\kappa^+ \delta x_+)]^2 -\frac {  [\text{Ai}'(\kappa^+ \delta x_+)]^2} {\kappa^+}\bigg]
\end{align}

and, similarly,
\begin{align}
   & \langle \hat\rho(x\approx x_-(t))\rangle= \frac{\theta_0(s_a)}{2\pi} -J_{ab} \frac{\cos\left(\Delta\Phi_t[x_0(s_a),x_0(s(x))]\right)}{\pi|s_a-s(x)|} \nn \\
   &-J_{ac} \frac{\sin\left(\Delta\Phi_t[x_0(s_a),x_0(s(x))]\right)}{\pi|s_a-s(x)|}\nn \\
    & + (\kappa^-)^2 \bigg[  \delta x_- [\text{Ai}(\kappa^- \delta x_-)]^2 -\frac {  [\text{Ai}'(\kappa^- \delta x_-)]^2} {\kappa^-}\bigg].
\end{align}


\section{Dressing relations in the low-coupling limit}\label{app:semi-classical}
This appendix contains some basic results that are used at small coupling in Sec. \ref{section:semi-classical_limit}. We remark that this small coupling limit is different from that of NLS at finite temperature considered in Ref.~\cite{del_Vecchio_del_Vecchio_2020} . Here, we are interested to zero temperature and weak coupling, which turns out to be a rather singular limit. An important formula, from which the dressing operation follows, is the so-called Tricomi formula \cite{tricomi1985integral}, as we now explain. Let us consider the principal value integral (or finite Hilbert transform)
\begin{equation}
    \fint_{a}^b \frac{f(x')}{x-x'} dx'   = g (x).
\end{equation}
The latter can be inverted using the following formula
\begin{equation}
    f(x)  = \frac{A - \pi^{-1} \int_a^b dt \frac{\sqrt{t - a} \sqrt{b-t}}{x-t} g(t) }{\pi \sqrt{(x-a)(b-x)}} ,
\end{equation}
with
\begin{equation}
    A = \int_{a}^{b} dx \ f(x) .
\end{equation}
However, it is easy to see that this constant is zero for the quantities below. In fact, this solution only gives the leading behaviour in $c$ when taking the low coupling limit of the Lieb-Liniger gas. The values at $\theta=\theta_F$, instead, require a different treatment. Specifying to the quasicondensate regime, the dressing operation  \eqref{eq:dressing_NLS-0} for $f(\theta) = 1$ gives
\begin{equation}
    1 = \int_{-\theta_F}^{\theta_F} \frac{d \mu}{2\pi} \frac{2c}{\lambda-\mu} \partial_\mu1^{\rm dr}(\mu),
\end{equation}
which is solved by
\begin{equation}\label{eq:semi_circle_condensate}
    \partial_\theta 1^{\rm dr}(\theta) = \frac{-\theta}{c \sqrt{\theta_F^2 - \theta^2}}.
\end{equation}
Once integrated, the last equation gives the semi-circle expression \eqref{eq:semicircle}. Recalling that the root density is $2\pi\rho(\theta)=1^{\rm dr}(\theta)$, we can then access all thermodynamic quantities in homogeneous states starting from \eqref{eq:semicircle}. For non-homogeneous states, one can use LDA and replace $\theta_F\to \theta_F(x)$. The effective velocity is obtained using $\varepsilon(\theta) = \theta^2/2$,   
\begin{equation}
    \theta = \int_{-\theta_F}^{\theta_F} \frac{d\mu}{2\pi} \frac{2 c}{\theta - \mu} \partial_\mu  [\theta^{\rm dr}(\mu)] ,
\end{equation}
which we invert to get 
\begin{equation}
   \theta^{\rm dr}(\theta)  = \frac{\theta\sqrt{\theta^2_F - \theta^2}}{2c} = v^{\rm eff}(\theta) 1^{\rm dr}(\theta).
\end{equation}
This equation can be solved for effective velocity. Furthermore, it also reproduces the static pressure $\mathcal{P}$ of a quasicondensate  
\begin{equation}
    \mathcal{P} = \int_{-\theta_F}^{\theta_F} d\theta \rho(\theta) v^{\rm eff}(\theta) \theta =\frac{\theta_F^4}{16c} = c\rho^2.
\end{equation}
\section{Proof of a useful identity}\label{app:great-identity}
In this appendix, we prove the useful identity
\begin{equation}\label{eq:the_great_identity}
 \pi \rho_{\rm LDA} =    p^{\rm Dr}_F = \theta^{\rm dr}(\theta_F) \sqrt{K},
\end{equation}
where here and in the following we drop the subscript ${}_{[n]}$ since the relation is independent of the filling $n$ with respect to which the dressing is done. 
To show this, we consider the well-known form of LDA density, and we perform integration by parts
\begin{eqnarray}
    \rho_{\rm LDA} &=& \int_{-\theta_F}^{\theta_F} \frac{d\theta}{2\pi} \dr 1(\theta) = \int_{-\theta_F}^{\theta_F} \frac{d\theta}{2\pi} (\partial_\theta \theta) \dr 1(\theta) \nn\\
    &=& \frac{ \theta_F \dr 1(\theta_F)}{\pi}- \int_{-\theta_F}^{\theta_F} \frac{d\theta}{2\pi}  \theta \partial_\theta \dr 1(\theta).
    \end{eqnarray}
Now an identity for derivatives of dressed functions $\partial_\theta \dr f(\theta)$ is inserted, which quickly simplifies the expression
    \begin{eqnarray}
      \rho_{\text{LDA}}  &=& \frac{ \theta_F \dr 1(\theta_F)}{\pi}+ \int_{-\theta_F}^{\theta_F} \frac{d\theta}{2\pi}  \theta (\dr T(\theta,\theta_F) \dr 1(\theta_F)) \nn\\ &&- \int_{-\theta_F}^{\theta_F} \frac{d\theta}{2\pi}  \theta \left(\dr T(\theta,-\theta_F) \dr 1(-\theta_F)\right)\nn\\
    &=& \frac{1}{2\pi }( \dr \theta(\theta_F) \dr 1(\theta_F) - \dr\theta(-\theta_F) \dr 1(-\theta_F)),\nn \\
\end{eqnarray}
having identified $\dr\theta(\pm\theta_F)$ with its definition. By recalling $\dr\theta(\theta_F) = -\dr \theta(-\theta_F)$ and $\dr 1(\theta_F) = \dr 1(-\theta_F)\equiv\sqrt{K}$, the identity in eq.~\eqref{eq:the_great_identity} follows.

\section{Saddle point corrections to LDA density in the low-coupling limit}\label{app:qghd-nls}
In this appendix, we determine the saddle point corrections to LDA density, corresponding to the quantum fluctuations, in the Gross-Pitaevskii limit discussed in Sec.~\ref{section:semi-classical_limit}. In doing so, we shall approximate the Luttinger parameter with its background value $K(x)\simeq K_0$ since this simplification does not change the scaling $K\sim 1/c$, and thus it does not change our conclusion below. \\

The calculation proceeds similarly to the Tonks-Girardeau case (see Sec.~\ref{sec:tonks} and Appendix~\ref{app:correction-lda-tg}), therefore we here limit to highlight the main differences. First, it is possible to define a coordinate along the initial Fermi surface (cf with eq.~\eqref{eq:s-coo})
\be
s(x_0)=\int_{-\infty}^{x_0} \frac{dx'}{v^{\rm eff}[\theta_F(x')]}=\int_{-\infty}^{x_0} \frac{dx'}{\sqrt{\rho_\text{NLS}(x')}},
\ee
where we used that $v^{\rm eff}(\theta)=\theta/2$ and eq.~\eqref{eq:nls-fermi-edges}. In terms of this coordinate, using that $n_\text{Fluct}(t;x,k)=n_\text{Fluct}(0; x-v^{\rm eff}(k)t,k)$, we can write the time-evolved density as
\be\label{eq:time-evo-dens-nls}
\langle\hat\rho(t,x)\rangle\simeq\sqrt{K_0}\iint \frac{dk dy}{2\pi^2} J_{12} \frac{\sin\left(\int_{x_1}^{x_2}\theta_F^{\rm dr}(x_t') \frac{dx_t'}{dx_0} dx_0\right)}{s_2-s_1} \ e^{iky}
\ee
with $J_{12}=\sqrt{|ds_1/dx_t||ds_2/dx_t|}$, and backward-evolved coordinates
\be
x_{1,2}\equiv x_0(s_{1,2})=x\mp y/2 - v^{\rm eff}(k) t.
\ee
Similarly to eq.~\eqref{eq:jacobians}, the jacobian is easily evaluated from the equation of motion $x_t(s)=x_0(s)+t\theta_0(s)/2$ and reads as
\be
\frac{dx_t(s)}{ds}=\sqrt{\rho_\text{NLS}[x_0(s)]}\left(1+\frac{t}{2} \frac{\rho_\text{NLS}'[x_0(s)]}{\sqrt{\rho_\text{NLS}[x_0(s)]}}\right).
\ee
Moreover, this allows us to rewrite the quantity $\theta^{\rm dr}_F(x_t) \frac{dx_t}{dx_0}  \equiv {\cal F}(x_0) $ entering in the phase of eq.~\eqref{eq:time-evo-dens-nls} as
\be
{\cal F}(x)=\sqrt{\frac{\pi}{c}} \rho_\text{NLS}^\frac{3}{4}(x)\left(1+\frac{t}{2}\frac{\rho_\text{NLS}'(x)}{\sqrt{\rho_\text{NLS}(x)}}\right).
\ee
At this point, similarly to the Tonks-Girardeau regime, the saddle point for the semi-classical action $S(k,y)=-ky + \int_{x_1}^{x_2} dx' \ {\cal F}(x')$ sets the main contribution of eq.~\eqref{eq:time-evo-dens-nls} to be
\be\label{eq:saddle-nls}
\langle\hat\rho(t,x)\rangle\simeq \sqrt{K_0}\sum_{\{s_{1,2}\}} J_{12} \frac{\sin\left(\int_{x_1}^{x_2} dx' {\cal F}(x')\right)}{2\pi(s_2-s_1)}
\ee
with the set $\{s_{1,2}\}$ specified by the solutions of 
\begin{align}
\label{eq:small_c_x}
x_{1,2}\equiv x_0(s_{1,2})&= x\mp y/2-v^{\rm eff}(\theta^{\rm dr}[x_0(s_{1,2})] )t\nn \\
&= x\mp y/2-t g[x_0(s_{1,2})]/\sqrt{c}, 
\end{align}
where in the second line we wrote $v^{\rm eff}(\theta^{\rm dr}(x_0))= g(x_0)/\sqrt{c}$ ($g$ a function of order ${\cal O}(1)$), since we expect that $v^{\rm eff}$ to have the same order of scaling as $\theta^{\rm dr}_F$.\\

It is then easy to see that the ``diagonal'' contributions in the sum of eq.~\eqref{eq:saddle-nls}, i.e. those for which $y\to0$ corresponds to $x_1\to  x_2$, reproduce the LDA density. For instance,
\be
\langle\delta\hat\rho\rangle\bigg\vert_{i,i}\simeq \sqrt{K_0}\frac{\theta^{\rm dr}_F(x_0(s_i))}{2\pi}=\frac{p^{\rm Dr}(x_0(s_i))}{2\pi}\sim{\cal O}(1/c).
\ee 
On the other hand, ``non-diagonal'' contributions $(i,j)$ give
\be
\langle\delta\hat\rho\rangle\bigg\vert_{i,j}\simeq \sqrt{K_0}J_{ij} \frac{\sin\left(\int_{x_i}^{x_j} dx' {\cal F}(x')\right)}{2\pi(s_i-s_j)}\sim {\cal O}(1).
\ee
At small coupling $c\to 0^+$ eq.~\eqref{eq:small_c_x} implies $x_j-x_i\sim{\cal O}(\sqrt{1/c})$ from which it follows that $s_j-s_i \sim{\cal O}(\sqrt{1/c})$. Thus, the phase $\int_{x_i}^{x_j} dx' {\cal F}(x')\sim {\cal O}(1/c)$ scales as the density, while the amplitude $\sim {\cal O}(1)$ since 
$\sqrt{K} \sim{\cal O}(1/\sqrt{c})$ and all other quantities are expressed in terms of rescaled variables. We can therefore conclude that quantum ripples in the small coupling limit $c\to0^+$ are highly-oscillating subleading corrections to $\rho_\text{LDA}\sim{\cal O}(1/c)$.
\bibliography{bibliography-new}

\end{document}